\documentclass[preprint2,tighten]{aastex6}
\pdfoutput=1 
\usepackage{amsmath,amstext}
\usepackage[T1]{fontenc}
\usepackage{apjfonts} 
\usepackage[figure,figure*]{hypcap}
\usepackage{multirow}
\usepackage[T1]{fontenc} 
\usepackage[utf8]{inputenc} 
\usepackage{float}

\newcommand{\unit}[1]{\ensuremath{\, \mathrm{#1}}}
\shorttitle{A High Mass and a Low Obliquity for the Young Neptune K2-25b}
\shortauthors{Stefansson et al. 2019}

\begin{document}
\title{The Habitable-zone Planet Finder Reveals A High Mass and a Low Obliquity for the Young Neptune K2-25b}
\author{Gudmundur Stefansson\altaffilmark{1,2,3,4,5}}
\author{Suvrath Mahadevan\altaffilmark{1,2}}
\author{Marissa Maney\altaffilmark{1}}
\author{Joe P. Ninan\altaffilmark{1,2}}
\author{Paul Robertson\altaffilmark{6}}
\author{Jayadev Rajagopal\altaffilmark{7}}
\author{Flynn Haase\altaffilmark{7}}
\author{Lori Allen\altaffilmark{7}}
\author{Eric B. Ford\altaffilmark{1,2,8}}
\author{Joshua Winn\altaffilmark{4}}
\author{Angie Wolfgang\altaffilmark{1,2}}
\author{Rebekah I. Dawson\altaffilmark{1,2,9}}
\author{John Wisniewski\altaffilmark{10}}
\author{Chad F. Bender\altaffilmark{11}}
\author{Caleb Ca\~nas\altaffilmark{1,2,3}}
\author{William Cochran\altaffilmark{12}}
\author{Scott A. Diddams\altaffilmark{13,14}}
\author{Connor Fredrick\altaffilmark{13,14}}
\author{Samuel Halverson\altaffilmark{15}}
\author{Fred Hearty\altaffilmark{1,2}}
\author{Leslie Hebb\altaffilmark{16}}
\author{Shubham Kanodia\altaffilmark{1,2}}
\author{Eric Levi\altaffilmark{1}}
\author{Andrew J. Metcalf\altaffilmark{17,13,14}}
\author{Andrew Monson\altaffilmark{1,2}}
\author{Lawrence Ramsey\altaffilmark{1,2}}
\author{Arpita Roy\altaffilmark{18,19}}
\author{Christian Schwab\altaffilmark{20}}
\author{Ryan Terrien\altaffilmark{21}}
\author{Jason T. Wright\altaffilmark{1,2}}
\email{gstefansson@astro.princeton.edu}

\altaffiltext{1}{Department of Astronomy \& Astrophysics, The Pennsylvania State University, 525 Davey Lab, University Park, PA 16802, USA}
\altaffiltext{2}{Center for Exoplanets \& Habitable Worlds, University Park, PA 16802, USA}
\altaffiltext{3}{NASA Earth and Space Science Fellow}
\altaffiltext{4}{Department of Astrophysical Sciences, Princeton University, 4 Ivy Lane, Princeton, NJ 08540, USA}
\altaffiltext{5}{Henry Norris Russell Fellow}
\altaffiltext{6}{Department of Physics and Astronomy, University of California-Irvine, Irvine, California 92697, USA}
\altaffiltext{7}{NSF's OIR Lab, 950 N. Cherry Ave. Tucson, AZ 85719, USA}
\altaffiltext{8}{Institute for Computational \& Data Sciences, The Pennsylvania State University, 203 Computer Building, University Park, PA 16802, USA}
\altaffiltext{9}{Alfred P. Sloan Foundation Fellow}
\altaffiltext{10}{Homer L. Dodge Department of Physics and Astronomy, University of Oklahoma, 440 W. Brooks Street, Norman, OK 73019, USA}
\altaffiltext{11}{Steward Observatory, University of Arizona, Tucson, Arizona 85721, USA}
\altaffiltext{12}{McDonald Observatory and Center for Planetary Systems Habitability, The University of Texas at Austin, Austin TX 78712, USA}
\altaffiltext{13}{Time and Frequency Division, National Institute of Standards and Technology, 325 Broadway, Boulder, CO 80305, USA}
\altaffiltext{14}{Department of Physics, University of Colorado, 2000 Colorado Avenue, Boulder, CO 80309, USA}
\altaffiltext{15}{Jet Propulsion Laboratory, California Institute of Technology, 4800 Oak Grove Drive, Pasadena, CA 91109, USA}
\altaffiltext{16}{Department of Physics, Hobart and William Smith Colleges, 300 Pulteney Street, Geneva, NY 14456, USA}
\altaffiltext{17}{Space Vehicles Directorate, Air Force Research Laboratory, 3550 Aberdeen Ave SE, Kirtland AFB, NM 87117, USA}
\altaffiltext{18}{Millikan Prize Postdoctoral Fellow}
\altaffiltext{19}{California Institute of Technology, 1200 E California Blvd, Pasadena, California 91125, USA}
\altaffiltext{20}{Department of Physics and Astronomy, Macquarie University, Balaclava Road, North Ryde, NSW 2109, Australia}
\altaffiltext{21}{Department of Physics and Astronomy, Carleton College, Northfield, Minnesota 55057, USA}

\begin{abstract}
Using radial-velocity data from the Habitable-zone Planet Finder, we have measured the mass of the Neptune-sized planet K2-25b, as well as the obliquity of its M4.5-dwarf host star in the 600-800MYr Hyades cluster. This is one of the youngest planetary systems for which both of these quantities have been measured, and one of the very few M dwarfs with a measured obliquity. Based on a joint analysis of the radial velocity data, time-series photometry from the {\it K2} mission, and new transit light curves obtained with diffuser-assisted photometry, the planet's radius and mass are $3.44\pm 0.12 \mathrm{R_\oplus}$ and $24.5_{-5.2}^{+5.7} \unit{M_\oplus}$. These properties are compatible with a rocky core enshrouded by a thin hydrogen-helium atmosphere (5\% by mass). We measure an orbital eccentricity of $e=0.43 \pm 0.05$. The sky-projected stellar obliquity is $\lambda=3 \pm 16^{\circ}$, compatible with spin-orbit alignment, in contrast to other "hot Neptunes" that have been studied around older stars.
\end{abstract}

\section{Introduction}
The observed orbital properties of planetary systems are influenced both by the formation process as well as subsequent dynamical interactions that can take place after planets are formed. Dynamical interactions over billions of years can modify or even rearrange planetary orbits, making it difficult to learn about the initial conditions. Young systems have had less time to evolve, and their orbital properties may provide a clearer view of the early stages of planet formation. As such, young systems are valuable benchmarks for testing models of planet formation and evolution.

The \textit{K2} mission \citep{howell2014} enabled the detection of many planets in young associations and clusters, including a number of planets discovered by The Zodical Exoplanets In Time (ZEIT) project \citep[e.g.,][]{mann2016,mann2016k233,mann2017}, and the four newborn transiting planets around V1298 Tau \citep{david2019}. The ongoing Transiting Exoplanet Survey Satellite (\textit{TESS}) mission \citep{Ricker2014} is also observing young stars, and has led to the discovery of the 45\,Myr old Neptune-sized planet DS Tuc Ab \citep{newton2019,benatti2019}, and the 22 MYr old Neptune orbiting the pre-main-sequence star AU Microscopii (AU Mic) \citep{plavchan2020}. Follow-up spectroscopic observations of both DS Tuc Ab \citep{zhou2019,montet2019}, and AU Mic b \citep{addison2020,hirano2010,palle2020} have revealed that both stars have low obliquities---the angle between its rotation axis and the planet's orbital axis. This is interesting because close-orbiting Neptunes around older stars seem to have a broad range of obliquities \citep[e.g.,][]{winn2010,bourrier2018}, although the number of measurements is still quite limited.

Both the stellar obliquity and orbital eccentricity are clues about the formation and subsequent dynamical history of planetary systems. For example, based on direct imaging data, \cite{bowler2019} reported a difference in the eccentricity distributions of planets and brown dwarfs, evidence that these objects form in different ways. Planets are expected to form on circular and coplanar orbits within protoplanetary disks, although can develop non-zero eccentricities via planet/planet interactions \citep{rasio1996}, secular von Zeipel-Lidov-Kozai cycles \citep{naoz2016,ito2019}, planet-disk interactions \citep{goldreich2003}, or other dynamical processes. The same dynamical processes can also alter orbital inclinations of planetary orbits, changing the obliquity of the star \citep{winn2015}. In particular, von Zeipel-Kozai-Lidov cycles combined with tidal friction \citep{fabrycky2007} can leave a planet stranded on a polar orbit \citep[e.g., GJ 436b;][]{bourrier2018} or even a retrograde orbit \citep[e.g., HAT-P-7b;][]{winn2009}. Obliquities can be measured by exploiting the Rossiter-McLaughlin (RM) effect, the alteration of the star's absorption line profiles during a planetary transit, which is often manifested as a radial-velocity anomaly \citep{rossiter1924,mclaughlin1924}.

Although hundreds of obliquities have been measured with the RM effect, the current list includes only three M dwarfs: GJ\,436 \citep{bourrier2018b}, TRAPPIST-1 \citep{hirano2020}, and AU Mic \citep{addison2020,hirano2020mic,palle2020}. The obliquity of GJ\,436b was found to be $\lambda = 72_{-24}^{+33\:\circ}$, suggesting a strong misalignment, and AU Mic was found to be well-aligned with the equator of its host star. For TRAPPIST-1, the current data are compatible with a low obliquity but with large uncertainties. RM measurements of M dwarfs have been limited as they tend to be optically faint, hindering the detection of the RM effect. If we could expand the sample of M-dwarfs with measured obliquities, we might be able to gain clues about the dynamical histories of late-type stars with close-in Neptune and Jupiter mass planets.

This paper reports on a suite of follow-up observations of K2-25, a young M4.5 dwarf in the Hyades with a close-orbiting and transiting Neptune-sized planet. The new data allow us to measure the planet's mass and the star's obliquity. Section \ref{sec:k225} introduces the K2-25 system. Section \ref{sec:observations} presents the new photometric and spectroscopic observations. Section \ref{sec:data} describes the data reduction. Section \ref{sec:stellarparam} presents an updated determination of the stellar parameters, along with new estimates of the projected rotation velocity, rotation period, and stellar inclination with respect to the line of sight. Section \ref{sec:params} presents a joint analysis of the photometric and spectroscopic data and provides the results for the planet's mass and other system parameters. Section \ref{sec:rm} focuses on the RM effect. All of the results are discussed in Section \ref{sec:discussion}, along with the feasibility of future observations of the planet's transmission spectrum. We conclude with a summary of our key findings in Section \ref{sec:summary}.

\section{The K2-25 System}
\label{sec:k225}
K2-25b was originally discovered by \cite{mann2016} and \cite{david2016} in data from the \textit{Kepler} spacecraft as part of the \textit{K2} mission. K2-25b is a Neptune-sized planet ($R \sim 3.5 R_\oplus$) in a $P = 3.5 \unit{day}$ orbit around its M4.5 dwarf host star in the Hyades. With its large transit depth of 1.1\% and its brightness at near-infrared (NIR) wavelengths ($J=11.3$), K2-25b has been discussed \citep[see e.g.,][]{mann2016,david2016} as a prime candidate for atmospheric characterization in the future with JWST, ARIEL \citep{tinetti2016}, and large ground-based observatories. In addition, due to its large transit depth and the rapid stellar rotation ($P=1.878 \unit{days}$), K2-25b has an estimated RM effect amplitude of $\sim60-70 \unit{m/s}$, making the Rossiter-McLaughlin effect detectable with high precision radial velocities in the NIR. Further, planets with well-constrained ages are scarce, making the determination of K2-25b's 3D orbit valuable for constraining theories of planet formation and migration mechanisms that aim to explain planetary and orbital parameters as a function of age.

Recently, two groups discussed additional transit follow-up observations of K2-25b. \cite{thao2020} study the transit depth of K2-25b as a function of wavelength using photometric observations at different wavelengths from \textit{K2}, the MEarth Observatories \citep{irwin2015}, and the Las Cumbres Global Telescope Network (LCOGT) in the optical, and from the \textit{Spitzer} Telescope in the NIR, and find no significant evidence of changes in transit depth as a function of wavelength. To explain the flat broadband transmission spectrum, they favor a scenario where K2-25b has a cloudy atmosphere using a predicted mass from exoplanet mass-radius relations. Although K2-25 could have a cloudy atmosphere, we show in this paper that the apparent flat broad-band transmission spectrum of K2-25b could also be partially explained by the larger as-observed mass of K2-25b than the mass assumed in \cite{thao2020}. 

\cite{kain2019} performed long-term photometric monitoring of K2-25 using the MEarth Observatories \citep{irwin2015}, to monitor the activity of the star, and to look for evidence of transiting timing variations (TTVs) in the transits of K2-25b, which could be suggestive of an additional planet orbiting in the system. They show that the photometric variability of the star was significantly smaller between 2016-2018 than what \textit{K2} observed in its 71 day observing window in 2015. \cite{kain2019} do not identify any definite starspot-crossing events in their transit data, although a few of their transits could contain tentative evidence for such events. They further searched for TTVs in the system from the additional transits, and find no evidence of significant TTVs, placing constraints on planetary companions orbiting close to mean-motion resonances of K2-25b's orbit. This agrees with the transit observations presented in this work.

\section{Observations}
\label{sec:observations}
\subsection{ARCTIC}
We obtained five transits of K2-25b using the Astrophysical Research Consortium Telescope Imaging Camera (ARCTIC) imager \citep{huehnerhoff2016} on the 3.5m Astrophysical Research Consortium (ARC) 3.5m Telescope at Apache Point Observatory (APO) on the nights of UT 20170917, 20190104, 20190118, 20190125, and 20190201. All of the transit observations were performed with the Engineered Diffuser available on ARCTIC, which we designed specifically to enable very high-precision photometric observations \citep[see e.g.,][]{stefansson2017,stefansson2018a,stefansson2018b,stefansson2020}. We used the SDSS $i^\prime$ filter except on the night of 20190118, when we used the SDSS $z^\prime$ filter without the diffuser to minimize background Moon contamination that night. Only data from the egress from this night were usable due to the moon contamination. The observations were performed in the quad-readout and $4\times4$ binning mode, resulting in a readout time of $2.7 \unit{s}$. The first transit was observed with an exposure time of $20 \unit{s}$, and the rest of the transits with an exposure time of $30 \unit{s}$. In this binning mode ARCTIC has a gain of $2.0 \unit{e/ADU}$, and a plate scale of $0.44 \unit{\arcsec/pixel}$. Table \ref{tab:obs} further summarizes the observations.

\subsection{Half Degree Imager}
We observed four transits of K2-25b using the Half-Degree Imager \citep[HDI;][]{hdi2013} at the WIYN 0.9m Telescope at Kitt Peak National Observatory on the nights of UT 20180207, 20180221, 20181214, and 20181221. HDI has a $4096 \times 4096 \unit{pixel}$ back-illuminated CCD from e2v, covering a $29.2\arcmin \times 29.2\arcmin$ Field-of-View (FOV) at a plate scale of $0.425 \unit{\arcsec/pixel}$, with a gain of $1.3 \unit{e / ADU}$ in the $1 \times 1$ binning mode. The observations were performed as part of commissioning observations of the newly installed Engineered Diffuser for the telescope, which is now available for high precision photometric observations. The Engineered Diffuser on HDI uses the same custom-optimized top-hat Engineered Diffuser pattern as we developed for the ARC 3.5m Telescope with a diffuser opening angle of $\theta = 0.34^{\circ}$ \citep{stefansson2017}. The diffuser is placed in a filter wheel holder in the dual filter wheel $45 \unit{mm}$ away from the focal plane, resulting in a stabilized Point Spread Function (PSF) with a Full-Width-at-Half-Maximum (FWHM) of $7.6 \unit{\arcsec}$. The size of the diffuser is $50.8 \times 50.8 \unit{mm}$ and vignettes the field to an effective FOV of $\sim20\arcmin \times 20\arcmin$, still allowing for a number of available reference stars in the field.

Table \ref{tab:obs} summarizes the observations, and lists the number of reference stars and exposure time used. All of the observations were performed in the SDSS $z^\prime$ filter using the $1 \times 1$ binning mode. The first three observations were performed with the Engineered Diffuser with an exposure time of 120s. The diffuser was not used on the night of 20181221 due to the high degree of Moon contamination (Moon was $\sim$97\% full and at a separation of $\sim6^{\circ}$ from the target). Further, during this night, we experienced issues with the camera shutter causing the shutter to be stuck for a periods of time which led to gaps in the datastream seen in Figure \ref{fig:transits}g. As the observations on that night were performed close to in-focus, the exposure time used was scaled down to 30s to minimize risk of saturation. Two different readout modes were used for the observations: a quad readout mode, and a single-amplifier readout mode, with readout times of $11 \unit{s}$ and $37 \unit{s}$, respectively. The observations on the nights of 20180207 and 20181221 used the single amplifier readout mode, with the quad readout mode used for the other two.

\subsection{Habitable-zone Planet Finder}
\label{sec:obshpf}
We obtained precision NIR RVs of K2-25 with the Habitable-zone Planet Finder (HPF) \citep{mahadevan2012,mahadevan2014} with the twofold goal to constrain the mass of K2-25b and to constrain the obliquity of the host star. HPF is a high resolution ($R\sim55,000$) NIR fiber-fed spectrograph on the 10m Hobby-Eberly Telescope (HET) at McDonald Observatory in Texas. HPF is actively temperature stabilized to the milli-Kelvin level to enable precision radial velocities in the NIR \citep{hearty2014,stefansson2016}. The HET is a fully queue-scheduled telescope with all observations executed in a queue by the HET resident astronomers \citep{shetrone2007}. HPF has a NIR laser-frequency comb (LFC) calibrator which has been shown to enable $\sim20\unit{cm\:s^{-1}}$ calibration precision and $1.5\unit{m\:s^{-1}}$ RV precision on-sky on the bright and stable M-dwarf Barnard's Star \citep{metcalf2019}.

In total, we obtained 105 spectra in 34 different visits with HPF. We removed 16 spectra after performing a S/N > 25 quality cut (S/N per pixel); these spectra were adversely affected by weather and deemed to be too low S/N. The median S/N of the 89 remaining spectra was 45, which were obtained in 32 different HET visits/tracks\footnote{The unique design of the HET only enables acquisition of K2-25 during a discrete East or West track with a $\sim$1 hour maximum track duration.}.

For the out-of-transit observations, we obtained 69 spectra of K2-25b, in 32 different tracks with HPF with a median S/N of 52 and median RV errorbar of $42 \unit{m/s}$. For most of the spectra (58 in total) the exposure time was $969 - 1160$s, with two spectra obtained per HET visit. In addition to these spectra, we also use 11 spectra which were obtained during the out-of-transit baseline for the three transit observations described below. These spectra had an exposure time of 309s and a median S/N of 32. To simplify the analysis and obtain the highest precision errorbar per RV visit, we performed a weighted average of the RVs of the out-of-transit RVs following the methodology described in \cite{metcalf2019,stefansson2020}. The final RVs used for the out-of-transit modeling are shown in Table \ref{tab:rvsout} in the Appendix.

To constrain the obliquity of the K2-25b system, we obtained spectra of K2-25 during three transits of K2-25b on the nights of 20181221 UT, 20181228 UT, and 20190104 UT. We used an exposure time of 309s, corresponding to 29 up-the-ramp reads on the HPF Hawaii-2RG NIR detector, to obtain the required time resolution to resolve the RM waveform. Overall, we obtained 11, 11, and 9 spectra in the three different visits, respectively. The three sets of spectra had a median S/N of 33, 36, and 35 (per extracted 1D pixel), and median RV errors of $85 \unit{m/s}$, $72 \unit{m/s}$, and $74 \unit{m/s}$, respectively. Out of these 31 spectra, 20 spectra were in transit and 11 out-of-transit. The S/N of the first night (20181221 UT), was the lowest as the Moon was $\sim$97\% full and only $\sim$$6^{\circ}$ away from the target during the observations. Making the S/N > 25 quality cut described above removed one out-of-transit baseline spectrum that was observed at the edge of the available HET track during the transit on 20190104 UT. Table \ref{tab:obs} further summarizes the in-transit observations. The final RVs used for the in-transit modeling are shown in Table \ref{tab:rvsin} in the Appendix.

Following our observational setup described in \cite{stefansson2020}, due to the faintness of the target ($J=11.3$), we elected to not have the HPF LFC on for any of the observations described, to minimize the risk that the bright LFC lines would contaminate the stellar spectrum. This does not significantly impact the drift correction for this target, as the drift of HPF is linear during a night (amplitude of $\sim$$10 \unit{m/s}$), and is well traced and calibrated to $<1\unit{m/s}$ RV precision by daily HPF calibrations and a linear drift model as has been detailed in \cite{stefansson2020}. For all of the observations above, the HPF sky fiber was used to subtract any Moon and/or other sources of background light contamination.

\begin{deluxetable*}{cccccccl}
\tablecaption{\label{tab:obs}Summary of transit observations analyzed in this work. The aperture setting column lists the aperture setting in pixels used to extract the photometry in AstroImageJ \citep{collins2017}, showing the radius of the photometric aperture, and the radius of the inner and outer background annulli, respectively. The Airmass Range column shows the airmass range of the observations from high to low airmass. The observations on UT 20181221 and UT 20190118 were not performed with the diffuser due to the large Moon contamination on those nights.}
\tablehead{\colhead{Date} & \colhead{Instrument} & \colhead{Filter/Bandpass} & \colhead{Exposure Time} & \colhead{Diffuser} & \colhead{Airmass Range} & \colhead{\# Ref. Stars} & \colhead{Aperture setting} \\
           \colhead{(UT)} & \colhead{}           & \colhead{}                & \colhead{(s)}           & \colhead{}         & \colhead{}              & \colhead{}             & \colhead{(pixels)}} 
\startdata
\multicolumn{7}{l}{\hspace{-0.4cm} Photometric Transit Observations:}           \\
20180207                  & HDI                  & SDSS z'                   & 120                     & Yes                & 2.10, 1.16             & 9                            & 21, 40, 60  \\ %
20180221                  & HDI                  & SDSS z'                   & 120                     & Yes                & 1.84, 1.09             & 12                           & 21, 40, 60  \\ %
20181214                  & HDI                  & SDSS z'                   & 120                     & Yes                & 1.70, 1.04             & 7                            & 22, 39, 59  \\ %
20181221                  & HDI                  & SDSS z'                   & 30                      & No                 & 1.37, 1.08             & 6                            & 9, 16, 24   \\ %
20170917                  & ARCTIC               & SDSS i'                   & 20                      & Yes                & 1.68, 1.09             & 8                            & 14, 35, 50  \\ %
20190104                  & ARCTIC               & SDSS i'                   & 30                      & Yes                & 1.41, 1.06             & 3                            & 16, 28, 42  \\
20190118                  & ARCTIC               & SDSS z'                   & 30                      & No                 & 1.23, 1.10             & 6                            & 16, 50, 70  \\
20190125                  & ARCTIC               & SDSS i'                   & 30                      & Yes                & 1.34, 1.05             & 13                           & 19, 30, 45  \\
20190201                  & ARCTIC               & SDSS i'                   & 30                      & Yes                & 1.11, 1.04             & 8                            & 16, 28, 42  \\
\multicolumn{7}{l}{\hspace{-0.4cm} Spectroscopic Transit Observations:}           \\
20181221                  & HPF                  & 820-1280nm                & 300                     & -                  & 1.37, 1.17             & -                            & -  \\
20181228                  & HPF                  & 820-1280nm                & 300                     & -                  & 1.36, 1.17             & -                            & -  \\
20190104                  & HPF                  & 820-1280nm                & 300                     & -                  & 1.34, 1.18             & -                            & -  \\
\enddata
\end{deluxetable*}

\section{Data Reduction} 
\label{sec:data}

\subsection{Photometric Observations}
We reduced the photometry using AstroImageJ \citep{collins2017}, following a similar methodology as in \cite{stefansson2017} and \cite{stefansson2018a}. In short, we experimented using a number of different aperture settings, varying the radii of the software aperture, and inner, and outer background annulli, selecting the aperture setting that resulted in the minimum RMS scatter in the resulting photometry. Table \ref{tab:obs} summarizes the aperture setting used for each observation in pixels that led to the lowest photometric noise in the light curve. We add the scintillation error bar estimates to the photometric error bars estimated by AstroImageJ following the methodology in \cite{stefansson2017}. The individual exposures were calibrated using standard median bias, dark, and flat field procedures in AstroImageJ following \cite{stefansson2017}. The median bias, dark, and flat-field exposures were taken at the beginning or end of each night of observation. We converted the timestamps of the ground-based observations to Barycentric Julian Date time ($\mathrm{BJD_{TDB}}$) using the Python package \texttt{barycorrpy} \citep{kanodia2018}, which uses the barycentric correction algorithm from \cite{wright2014}.

K2-25b was observed by the \textit{K2} mission as part of Campaign 4 in long-cadence (30min exposures) mode, from February 8, 2015 to April 20th 2015, resulting in 71 days of continuous photometric observations. We use the light curve from the \texttt{Everest} pipeline \citep{luger2016}, which is capable of correcting the periodic correlated errors in the \textit{K2} data due to imperfect pointing of the spacecraft. The corrected \texttt{Everest} light curve improved the $6 \unit{hour}$ CDPP standard deviation of the raw \textit{K2} long-cadence data of K2-25 from $432 \unit{ppm}$ to $295 \unit{ppm}$.

\subsection{Spectroscopic Observations}
The HPF 1D spectra were reduced and extracted with the custom HPF data-extraction pipeline following the procedures outlined in \cite{ninan2018}, \cite{kaplan2018}, and \cite{metcalf2019}. For the RV extractions, we used the Spectrum Radial Velocity Analyzer (SERVAL) which we have adapted for use for the HPF spectra, following the methodology described in \cite{metcalf2019} and \cite{stefansson2020}. In short, SERVAL uses the template-matching method to measure the radial velocities \cite[see e.g.,][]{anglada2012,zechmeister2018}. We extracted the HPF RVs using the 8 orders with the least telluric contamination in the HPF bandpass (orders covering the wavelength regions from 8540-8890\AA, and 9940-10760\AA). Although we plan to include more orders in the RV extraction to potentially enable even high RV precision, we elected using only these 8 orders that we have extensively tested to produce reliable RVs with HPF. We subtracted the estimated sky-background from the stellar spectrum using the dedicated HPF sky fiber. Following the methodology described in \cite{metcalf2019}, we explicitly masked out telluric lines and sky-emission lines to minimize their impact on the RV determination. To minimize the impact of the asymmetric spectral-line variations seen during the RM effect, we generated the master RV template to extract the RV using only out-of-transit spectra. As described in Subsection \ref{sec:obshpf}, this resulted in 40 spectra after a quality cut of S/N > 25 used to generate the master RV template.

We calculated the RVs in two steps. First, to generate a first-pass master template, we ignored any possible planetary induced RVs and coadded the out-of-transit spectra (after barycentric correction) using the \texttt{SERVAL} template creation functionality, which we then used to derive a first set of RVs. We then used this first set of RVs to further align the out-of-transit spectra to create a second-pass RV template, and re-extracted RVs using this more refined template. The out-of-transit RVs used for the mass determination of K2-25b are shown in Table \ref{tab:rvsout} in the Appendix, and the in-transit RVs are shown in Table \ref{tab:rvsin}.

\section{Stellar Parameters}
\label{sec:stellarparam}
The stellar parameters used in this work are summarized in Table \ref{tab:stellarparam}, and are broadly adopted from the values presented in \cite{mann2016} and \cite{thao2020}. \cite{thao2020} estimate their stellar parameters (including the stellar radius and mass) using the empirically calibrated relations of \cite{mann2015} and \cite{mann2019}. As an additional test of these parameters, we performed an independent SED and isochrone fit using the \texttt{EXOFASTv2} package \citep{eastman2017,eastman2019} placing informative Gaussian priors on the known metallicity ($\mathrm{[Fe/H]}=0.15 \pm 0.03$) and age $\mathrm{Age} = 730 \pm50 \unit{Myr}$ of the star derived from its Hyades cluster membership \citep{mann2016}, along with its known distance from \textit{Gaia}. The values we obtained are in good agreement with the values from \cite{thao2020}---in particular obtaining a fully consistent stellar density value from the stellar mass and radius (we obtain a density of $\rho = 14.7 \pm 1.2 \unit{g\:cm^{-3}}$, and they obtain $\rho = 14.7 \pm 1.5 \unit{g\:cm^{-3}}$), which carries important information on the eccentricity of the planet via the photoeccentric effect \citep[see e.g.,][]{dawson2012}. We elected to incorporate the values from \cite{thao2020} for our analysis, as they are derived from precise empirically calibrated relations, rather than through model-dependent SED and isochrone fitting. We note that with an effective temperature of $T_{\mathrm{eff}}=3207 \unit{K}$ and radius of $R=0.29R_\oplus$, K2-25 lies at the higher radius end of the radius discontinuity for low-mass M-dwarfs in \cite{rabus2019}, which they observe for M-dwarfs with effective temperatures between 3200K and 3300K. This radius discontinuity has been interpreted as the boundary between partially convective and fully-convective M-dwarfs, and as K2-25 is observed to be just above the radius discontinuity, suggests that K2-25 is potentially partially convective and just at the onset of being a fully-convective star. Our estimate of the stellar rotation period, projected rotational velocity, and stellar inclination are discussed in the next two subsections.

\begin{deluxetable}{llcc}
\tablecaption{Summary of stellar parameters used in this work. \label{tab:stellarparam}}
\tabletypesize{\scriptsize}
\tablehead{\colhead{~~~Parameter}                                 &  \colhead{Description}                                            & \colhead{Value}                         & \colhead{Reference}}
\startdata
\multicolumn{4}{l}{\hspace{-0.2cm} Main identifiers:}                                                                                                                                  \\
EPIC                                                              &  -                                                                & 210490365                               & Huber                  \\
\multicolumn{4}{l}{\hspace{-0.2cm} Stellar magnitudes:}           \\
$B$                                                               &  APASS Johnson B mag                                              & $17.449 \pm 0.144$                      & APASS                  \\
$V$                                                               &  APASS Johnson V mag                                              & $15.891 \pm 0.180$                      & APASS                  \\
$g^{\prime}$                                                      &  APASS Sloan $g^{\prime}$ mag                                     & $16.567 \pm 0.018$                      & APASS                  \\
$r^{\prime}$                                                      &  APASS Sloan $r^{\prime}$ mag                                     & $15.300 \pm 0.183$                      & APASS                  \\
$i^{\prime}$                                                      &  APASS Sloan $i^{\prime}$ mag                                     & $13.698 \pm 0.206$                      & APASS                  \\ 
\textit{Kepler}-mag                                               &  \textit{Kepler} magnitude                                        & $14.528$                                & Huber                  \\
$J$                                                               &  2MASS $J$ mag                                                    & $11.303\pm 0.021$                       & 2MASS                  \\
$H$                                                               &  2MASS $H$ mag                                                    & $10.732\pm 0.020$                       & 2MASS                  \\
$K_S$                                                             &  2MASS $K_S$ mag                                                  & $10.444\pm0.019$                        & 2MASS                  \\
$WISE1$                                                           &  WISE1 mag                                                        & $10.275\pm0.024$                        & WISE                   \\
$WISE2$                                                           &  WISE2 mag                                                        & $10.086\pm0.020$                        & WISE                   \\
$WISE3$                                                           &  WISE3 mag                                                        & $9.936\pm0.057$                         & WISE                   \\
\multicolumn{4}{l}{\hspace{-0.2cm} Stellar Parameters:}\\
$M_*$                                                             &  Mass in $M_{\odot}$                                              & $0.2634 \pm 0.0077$                     & \cite{thao2020}        \\
$R_*$                                                             &  Radius in $R_{\odot}$                                            & $0.2932 \pm 0.0093$                     & \cite{thao2020}        \\
$\rho_*$                                                          &  Density in $\unit{g\:cm^{-3}}$                                   & $14.7 \pm 1.5$                          & \cite{thao2020}        \\
$\log(g)$                                                         &  Surface gravity in cgs units                                     & $4.944 \pm 0.031$                       & Derived from $M$ \& $R$\\
$T_{\mathrm{eff}}$                                                &  Effective temperature in \unit{K}                                & $3207 \pm 58$                           & \cite{thao2020}        \\
$\mathrm{[Fe/H]}$                                                 &  Metallicity in dex                                               & $0.15 \pm 0.03$                         & \cite{mann2016}        \\
Age                                                               &  Age in Gyrs                                                      & $0.730_{-0.052}^{+0.050}$               & \cite{mann2016}        \\
$L_*$                                                             &  Luminosity in $L_\odot$                                          & $0.00816 \pm 0.00029$                   & \cite{thao2020}        \\
$d$                                                               &  Distance in pc                                                   & $45.01_{-0.17}^{+0.16}$                 & Gaia                   \\
$\pi$                                                             &  Parallax in mas                                                  & $22.218_{-0.083}^{+0.081}$              & Gaia                   \\
$P_{\mathrm{rot}}$                                                &  Rotational period in days                                        & $1.878 \pm 0.005$                       & This work              \\
$v \sin i_*$                                                      &  \multirow{2}{2.3cm}{Stellar rotational \hspace{0.3cm}\: velocity in \unit{km\ s^{-1}}}  & $8.8 \pm 0.6$    & This work              \\
                                                                  &                                                                   &                                         &                        \\
$i_*$                                                             &  Stellar inclination                                              & $90 \pm 12^{\circ}$                     & This work              \\
\enddata
\tablenotetext{}{References are: Huber \citep{huber2016}, Lepine \citep{lepine2005}, Reid \citep{reid2004}, Gaia \citep{gaia2018}, APASS \citep{henden2015apass}, UCAC2 \citep{zacharias2004ucac2}, 2MASS \citep{cutri20032mass}, WISE \citep{cutri2014wise}.}
\end{deluxetable}

\subsection{Rotation Period}
The \textit{K2} data of K2-25b is modulated at the $1\%$ level, suggestive of starspots and/or active regions that rotate in and out-of-view with the rotation period of the star. We independently measure the rotation period of K2-25b following the methodology in \cite{stefansson2020}. In short, we model the stellar active regions using a quasi-periodic Gaussian Process (GP) as quasi-periodic GP kernels have been shown to yield reliable stellar rotation rates \citep{Angus2018}. For computational efficiency, we adopt the quasi-periodic kernel from the \texttt{celerite} package \citep{Foreman-Mackey2017} as implemented in \texttt{juliet}. The form of this kernel is further discussed in Section \ref{sec:transitrvmodel}. To estimate the rotation period, we removed points in a 2x transit window around the expected transit times, and fit the resulting photometry using the \texttt{celerite} quasi-periodic GP kernel \citep[see][and Equation \ref{eq:kernelperiodic} in Subsection \ref{sec:transitrvmodel}]{Foreman-Mackey2017}. We placed non-informative priors on the GP period, amplitude and GP decay timescale parameters. 

\begin{figure}[t!]
\begin{center}
\includegraphics[width=1.0\columnwidth]{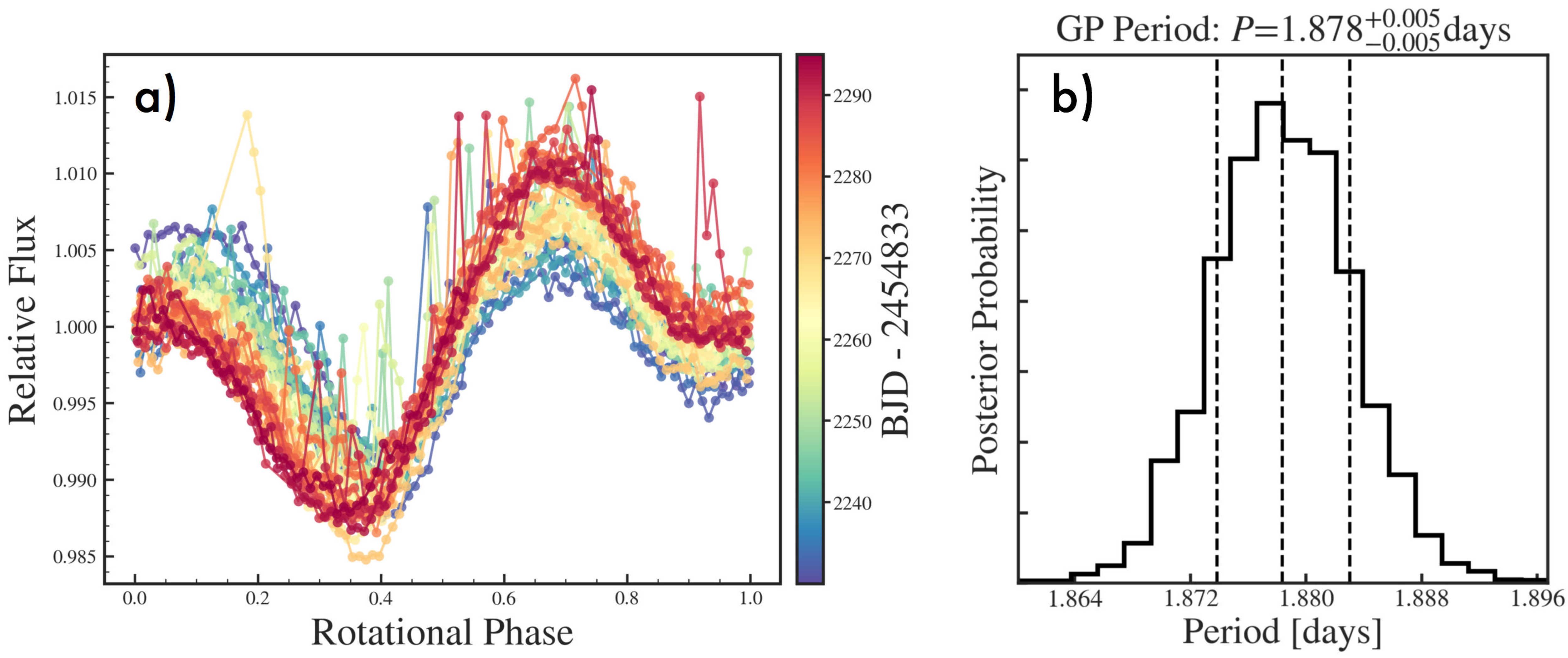}
\vspace{-0.5cm}
\caption{a) Phase-folded \textit{K2} photometry of K2-25 using our best estimate of the rotation period (see b), showing a clear periodic photometric modulation. The photometric modulation remains stable over the 71 day \textit{K2} photometric baseline, with a slight evolution seen. b) The posteriors of the period of the quasi-periodic GP kernel, which we interpret as the stellar rotation period.}
\label{fig:rotation}
\end{center}
\end{figure}

Figure \ref{fig:rotation} shows the resulting posteriors of the GP period hyperparameter along with the phase-folded photometry using our best-fit rotation period of $P_{\mathrm{rot}} = 1.878 \pm 0.005 \unit{days}$. In the phase-folded photometry, we see that the photometric modulation remains relatively stable throughout the 71 days of \textit{K2} dataset, with a slight evolution observed. From Figure \ref{fig:rotation} we also see evidence of flares. Our period estimate agrees well with the rotation period reported in \cite{mann2016} of $P_{\mathrm{rot}} = 1.881 \pm 0.021 \unit{days}$ estimated using an Autocorrelation-Function method, and \cite{dmitrienko2017} of $P_{\mathrm{rot}} = 1.878 \pm 0.030 \unit{days}$ from power-spectra analysis. The top panel in Figure \ref{fig:transits} in Section \ref{sec:params} shows the modulation in the \textit{K2} photometry as a function of time for the full 71 day baseline.

\subsection{Projected Rotational Velocity and Stellar Inclination}
We measured the projected rotation velocity using the empirical spectral matching algorithm described in \cite{stefansson2020}, which closely follows the \texttt{SpecMatchEmp} algorithm described in \cite{yee2017}. In short, the algorithm compares the as-observed target star spectrum to a library of as-observed slowly rotating stellar spectra using a $\chi^2$ metric, and refer the reader to \cite{stefansson2020} for a more detailed discussion.

\begin{table}[b!]
\centering
\caption{Resulting $v \sin i_*$ values for the 8 different HPF orders cleanest of tellurics. The resulting median value is $v \sin i_* = 8.8 \unit{km/s}$ with a scatter of $0.3 \unit{km/s}$. As is mentioned in the text, we adopt a value of $v \sin i_* = 8.8 \pm 0.6 \unit{km/s}$.}
\begin{tabular}{ccc}
\hline\hline
Order  &  Wavelength Region [\AA]  & $v \sin i_*$ [km/s] \\ 
\hline
4  & [8540, 8640]   & 8.3 \\
5  & [8670, 8750]   & 8.9 \\
6  & [8790, 8885]   & 8.6 \\
14 & [9940, 10055]  & 8.8 \\
15 & [10105, 10220] & 8.8 \\
16 & [10280, 10395] & 9.0 \\
17 & [10460, 10570] & 9.3 \\
18 & [10640, 10760] & 8.9 \\
\hline
\end{tabular}
\label{tab:vsini}
\end{table}

We used the algorithm to measure independent $v \sin i_*$ values for the 8 HPF orders cleanest of tellurics. Table \ref{tab:vsini} shows the resulting values, showing that the independent orders agree well on the resulting value, with a mean value of $v \sin i_* = 8.8 \unit{km/s}$ and a scatter of $0.3 \unit{km/s}$. The values in Table \ref{tab:vsini} show a small formal scatter at the $0.3 \unit{km/s}$ level. However, given our experience calculating $v \sin i_*$ values of other stars from high resolution spectra with HPF a more realistic error estimate is a factor of 2 larger. We adopt an error estimate of $0.6 \unit{km/s}$ to account for possible systematics in our method as $v \sin i_*$ measurements are generally dominated by systematics \citep[see e.g.,][]{reiners2012}. As such, we adopt a $v \sin i_* = 8.8 \pm 0.6 \unit{km/s}$. We note that value is somewhat larger than the value presented in \cite{mann2016} of $v \sin i_* = 7.8 \pm 0.5 \unit{km/s}$ determined from their as-observed spectra obtained with the IGRINS spectrograph. The method used in \cite{mann2016} used an overall similar $\chi^2$ method as presented here, but used artificially rotationally broadened theoretical BT-SETTL spectra for the $\chi^2$ comparison instead of as-observed spectra. Although this value is formally slightly higher than the maximum equatorial velocity of the star of $7.9 \pm 0.25 \unit{km/s}$ (assuming a stellar radius of $R = 0.2932 \pm 0.0093 \unit{R_\odot}$ and stellar rotation period of $P = 1.878 \pm 0.005 \unit{days}$) the two values overlap within the $2 \sigma$ uncertainties. The high $v \sin i_*$ suggests that the stellar inclination is close to $90^\circ$.

To estimate accurate posteriors for the stellar inclination $i_*$ from the stellar rotational velocity $v$ estimated from $R$ and $P_{\mathrm{rot}}$ and its sky-projection $v \sin i_*$ measured from the HPF spectra, we use the formalism in \cite{masuda2020}, which accurately accounts for the correlated dependence between $v\sin i_*$ and $v$. The $v\sin i_*$ measurement does not distinguish between solutions between $i$ and $180^{\circ}-i$ and we thus calculate two independent solutions between $0^{\circ}$ and $90^{\circ}$, and $90^{\circ}$ and $180^{\circ}$, respectively. Using the values listed in Table \ref{tab:stellarparam} for $P_{\mathrm{rot}}$, $v \sin i_*$, and $R$, we obtain two mirrored posteriors with a highest-likelihood inclination at $90^{\circ}$. Taken together, the two solutions result in an inclination constraint of $90 \pm 12^{\circ}$, consistent with viewing K2-25's stellar equator edge on. This agrees well with the stellar inclination constraint provided by \cite{mann2016} of $i_* > 79^\circ$ at 1$\sigma$ (68.4\% confidence; they considered inclinations between 0 and $90^\circ$).

\begin{deluxetable*}{llccc}
\tablecaption{Summary of priors used for the three joint transit and RV fits performed. $\mathcal{N}(m,\sigma)$ denotes a normal prior with mean $m$, and standard deviation $\sigma$; $\mathcal{U}(a,b)$ denotes a uniform prior with a start value $a$ and end value $b$, $\mathcal{J}(a,b)$ denotes a Jeffreys prior with a start value $a$ and end value $b$. A Gaussian prior on the stellar density was placed for all fits. The dilution parameters in \texttt{juliet} were fixed to 1 for all transit observations. For the three models considered we sampled 70, 72, and 69 parameters. \label{tab:priors1}}
\tablehead{\colhead{Parameter}&  \colhead{Description}                             & \colhead{Model RV1}                       & \colhead{Model RV2}                     & \colhead{Model RV3} \\
                              &                                                    & \colhead{($e=0$, GP for RVs)}           & \colhead{($e\neq0$, GP for RVs)}        & \colhead{($e\neq0$, No GP for RVs)}} 
\startdata	
\hline
\multicolumn{5}{l}{\hspace{-0.3cm} Orbital Parameters:}           \\                    
$P$ (days)                    &  Orbital period                                    & $\mathcal{N}(3.484548,0.000042)$        & $\mathcal{N}(3.484548,0.000042)$      & $\mathcal{N}(3.484548,0.000042)$       \\ 
$T_C$                         &  Transit Midpoint - 2400000 $(\mathrm{BJD_{TDB}})$ & $\mathcal{U}(58515.63,58515.66)$        & $\mathcal{U}(58515.63,58515.66)$      & $\mathcal{U}(58515.63,58515.66)$       \\ 
$r_1$$^a$                     &  Radius ratio / Impact parameter                   & $\mathcal{U}(0,1)$                      & $\mathcal{U}(0,1)$          		     & $\mathcal{U}(0,1)$                     \\
$r_2$$^a$                     &  Radius ratio / Impact parameter                   & $\mathcal{U}(0,1)$            	         & $\mathcal{U}(0,1)$                    & $\mathcal{U}(0,1)$                     \\ 
$a/R_*$                       &  Scaled semi-major axis                            & $\mathcal{U}(1,50)$           	         & $\mathcal{U}(1,50)$                   & $\mathcal{U}(1,50)$                    \\ 
$m_{\mathrm{flux}}$           &  Transit baseline parameter                        & $\mathcal{U}(0,0.1)$         	         & $\mathcal{U}(0,0.1)$                  & $\mathcal{U}(0,0.1)$                   \\ 
$\sigma_{\mathrm{K2}}$        &  Photometric errorbar (ppm)                        & $\mathcal{J}(1,1000)$                   & $\mathcal{J}(1,1000)$                 & $\mathcal{J}(1,1000)$                  \\ 
e                             &  Eccentricity                                      & 0 (fixed)                    	         & $\mathcal{U}(0,0.95)$                 & $\mathcal{U}(0,0.95)$                  \\ 
$\omega$                      &  Argument of Periastron                            & 90 (fixed)                              & $\mathcal{U}(0,360)$                  & $\mathcal{U}(0,360)$                   \\ 
$K$                           &  RV semi-amplitude ($\unit{m/s}$)                  & $\mathcal{U}(0,200)$                    & $\mathcal{U}(0,200)$                  & $\mathcal{U}(0,200)$                   \\ 
\multicolumn{5}{l}{\hspace{-0.3cm} Other constraints:}           \\
$\rho_*$                      &  Stellar density ($\unit{g\:cm^{-3}}$)             & $\mathcal{N}(14.7,1.5)$                 & $\mathcal{N}(14.7,1.5)$               & $\mathcal{N}(14.7,1.5)$                \\ 
\multicolumn{5}{l}{\hspace{-0.3cm} Jitter and other instrumental terms:}           \\
$q_1$$^{b}$                   &  Limb-darkening parameter                          & $\mathcal{U}(0,1)$            	         & $\mathcal{U}(0,1)$                    & $\mathcal{U}(0,1)$                     \\ 
$q_2$$^{b}$                   &  Limb-darkening parameter                          & $\mathcal{U}(0,1)$            	         & $\mathcal{U}(0,1)$                    & $\mathcal{U}(0,1)$                     \\ 
$\sigma_{\mathrm{phot}}$$^c$  &  Photometric jitter ($\unit{ppm}$)                 & $\mathcal{J}(1,1000)$                   & $\mathcal{J}(1,1000)$                 & $\mathcal{J}(1,1000)$                  \\ 
$\mu_{\mathrm{phot}}$$^c$     &  Photometric baseline                              & $\mathcal{N}(0,0.1)$                    & $\mathcal{N}(0,0.1)$                  & $\mathcal{N}(0,0.1)$                   \\ 
$\sigma_{\mathrm{HPF}}$       &  HPF RV jitter ($\unit{m/s}$)                      & $\mathcal{J}(0.1,300)$                  & $\mathcal{J}(0.1,300)$                & $\mathcal{J}(0.1,300)$                 \\ 
$\mu_{\mathrm{HPF}}$          &  HPF RV offset ($\unit{m/s}$)                      & $\mathcal{U}(-200,200)$                 & $\mathcal{U}(-200,200)$               & $\mathcal{U}(-200,200)$                \\ 
\multicolumn{5}{l}{\hspace{-0.3cm} Shared Photometric and RV Quasi-Periodic GP Parameters:}           \\
$P_{\mathrm{GP}}$             &  GP Period (days)                                  & $\mathcal{N}(1.8784,0.005)$             & $\mathcal{N}(1.8784,0.005)$           & $\mathcal{N}(1.8784,0.005)$            \\ 
\multicolumn{5}{l}{\hspace{-0.3cm} \textit{K2} Quasi-Periodic GP Parameters:}           \\
$B$                           &  Photometric GP Amplitude ($\unit{ppm^2}$)           & $\mathcal{J}(10^{-6},10^{5})$           & $\mathcal{J}(10^{-6},10^{5})$         & $\mathcal{J}(10^{-6},10^{5})$          \\ 
$C$                           &  GP Additive Factor                                & $\mathcal{J}(10^{-6},10^{5})$           & $\mathcal{J}(10^{-6},10^{5})$         & $\mathcal{J}(10^{-6},10^{5})$          \\ 
$L$                           &  GP Length Scale ($\unit{days}$)                   & $\mathcal{J}(10^{3},10^{6})$         	 & $\mathcal{J}(10^{3},10^{6})$          & $\mathcal{J}(10^{3},10^{6})$           \\ 
\multicolumn{5}{l}{\hspace{-0.3cm} Ground-based Approximate Matern GP Parameters$^{d}$:}           \\
$\sigma_{\mathrm{GP}}$        &  Photometric GP Amplitude ($\unit{ppm}$)           & $\mathcal{J}(10^{-1},10^{4})$           & $\mathcal{J}(10^{-1},10^{4})$         & $\mathcal{J}(10^{-1},10^{4})$          \\ 
$L$                           &  Timescale of exponential kernel ($\unit{days}$)   & $\mathcal{J}(10^{-2},10^{5})$         	 & $\mathcal{J}(10^{-2},10^{5})$         & $\mathcal{J}(10^{-2},10^{5})$          \\ 
$\rho$                        &  Timescale of Matern kernel (days)                 & $\mathcal{J}(10^{-2},10^{5})$           & $\mathcal{J}(10^{-2},10^{5})$         & $\mathcal{J}(10^{-2},10^{5})$          \\ 
\multicolumn{5}{l}{\hspace{-0.3cm} RV GP Parameters:}           \\
$\sigma_{\mathrm{GP}}$        &  RV GP Amplitude ($\unit{m/s}$)                    & $\mathcal{J}(10^{1},10^{5})$            & $\mathcal{U}(10^{1},10^{5})$          & -                                      \\ 
$\Gamma$                      &  Harmonic structure / scaling parameter            & $\mathcal{N}(8.0,1.9)$                  & $\mathcal{N}(8.0,1.9)$                & -                                      \\ 
$\alpha$                      &  Inverse length scale  ($\unit{days^{-2}}$)        & $\mathcal{J}(10^{-12},10^{-3})$     	 & $\mathcal{J}(10^{-12},10^{-3})$       & -                                      \\ 
\enddata
\tablenotetext{a}{Using the efficient sampling of the $r_1$ and $r_2$ parameterization for the impact parameter $b$ and radius ratio $p=R_p/R_*$ as described in \cite{espinoza2018b}.}
\tablenotetext{b}{Overall we modeled 4 pairs of limb darkening parameters $q_1$ and $q_2$ (parametrization from \cite{kipping2013}): a) one pair for \textit{K2}, b) one pair for all of the transits observed with ARCTIC in the SDSS i$^\prime$ filter and c) one pair for the ARCTIC SDSS z$^\prime$ filter observations, and d) one pair for the HDI transits in the SDSS z$^\prime$ filter.}
\tablenotetext{c}{We placed a separate photometric jitter term and baseline offset term for each of the photometric observations.}
\tablenotetext{d}{We place one set of three parameters ($\sigma_{\mathrm{GP}}$, $L$, $\rho$) for each of the ground-based transits.}
\end{deluxetable*}

\section{Planet Parameters from Transit Photometry and RVs}
\label{sec:params}

\subsection{Transit, RV and Gaussian Process model}
\label{sec:transitrvmodel}
We jointly model the \textit{K2} and ground-based transits and the HPF out-of-transit RVs using the \texttt{juliet} Python package \citep{Espinoza2019}, which uses the \texttt{batman} Python package \citep{kreidberg2015} for the transit model, and the \texttt{radvel} package \citep{fulton2018} for the RV model. We used the \texttt{dynesty} sampler \citep{speagle2019} available in \texttt{juliet} to perform dynamic nested sampling to obtain both posterior and evidence estimates, where we used the default weight and stopping functions, and stopping criteria in \texttt{dynesty}\footnote{For a description of the weight and stopping functions in the dynamic nested sampler in \texttt{dynesty}, see Section 3 in \cite{speagle2019}.}. The total log likelihood in \texttt{juliet} is the sum of the individual log-likelihoods of each dataset considered\footnote{See Equations and discussion surrounding Equations 6 and 7 in \cite{Espinoza2019} to see the explicit likelihood used in \texttt{juliet}.}.

Following the implementation in \texttt{juliet}, we parametrize the radius ratio $p = \mathrm{R_p/R_*}$ and the impact parameter $b$ using the efficient $r_1$ and $r_2$ parametrization described in \cite{espinoza2018b}. We sampled the limb darkening parameters using the quadratic $q_1$ and $q_2$ limb-darkening parametrization from \cite{kipping2013}. Following the suggestion in \cite{kipping2010}, we resampled and rebinned our transit model to the effective 30 minute exposure time of the long-cadence \textit{K2} data to account for the smoothing of the transit shape. We assumed that there was no source of dilution, as no obvious close-by companion is seen in the adaptive optics imaging presented in \cite{mann2016}. 

We used a Gaussian Process (GP) model with 3 different kernels to account for the characteristic correlated noise behavior in the \textit{K2} photometry, ground-based photometry, and the radial velocities. GPs have been used by a number of groups to jointly model correlated noise observed in photometric and radial velocity data due to stellar active regions on the surface of the star \citep[e.g., spots, faculae, plages; see e.g.,][]{haywood2014,grunblatt2015,lopezmorales2016,dai2017,Angus2018,haywood2018}. For our GP modeling, we use the GP kernels available in \texttt{juliet}, which are based on the GP implementations from the \texttt{george} \citep{ambikasaran2015} and \texttt{celerite} \citep{Foreman-Mackey2017} Python packages. In \texttt{juliet}, the elements of the covariance matrix $C_i$ for instrument $i$ are assumed to be of the form,
\begin{equation}
C_{i,l,m} = k_i(x_l,x_m) + (\sigma_{i,w}^2 + \sigma_{t_{l},i}^2)\delta_{l,m},
\label{eq:cov}
\end{equation}
where $\delta_{l,m}$ is the Kronecker delta function, $k_i(x_l,x_m)$ is the kernel of the GP for instrument $i$, and $\sigma_{t_l,i}$ is the error estimated at time $t_l$, and $\sigma_{i,w}$ is an additional white-noise jitter parameter. For all of the GP fits considered, we fit a kernel function as well as a separate white noise jitter term for each instrument as is shown in Equation \ref{eq:cov}. 

We choose three different GP kernels as a balance between computational speed and to adequately account for the characteristic correlated noise properties of the different datasets. First, to model the correlated noise in the radial velocity observations, we elect to use the quasi-periodic rotational kernel, which has been shown to be effective at modeling rotational variations in RV datasets and has well-studied hyperparameters \citep{haywood2014,grunblatt2015,haywood2018}. In \texttt{juliet}, the quasi-periodic kernel is given with the following form, 
\begin{equation}
k(x_l,x_m) = \sigma_{\mathrm{GP}}^2 \exp \left( - \alpha \tau^2 - \Gamma \sin^2 \left[ \frac{\pi \tau}{P_{\mathrm{GP}}}\right] \right),
\label{eq:kernelgeorge}
\end{equation}
where $\tau = |x_l - x_m|$, with hyperparameters $\sigma_{\mathrm{GP}}$ (RV amplitude in $\mathrm{m/s}$), $\alpha$ (inverse square timescale in units of $\unit{days^{-2}}$), $\Gamma$ (a unitless scaling parameter), and $P_{\mathrm{GP}}$ (the periodicity of the GP in days, which we interpret as the rotation period)\footnote{As mentioned by \cite{Espinoza2019}, the hyperparameters in Equation \ref{eq:kernelgeorge} correspond to the following parameters in the notation of \cite{haywood2018}: $\sigma_{\mathrm{GP}} = \eta_1$, $\alpha = 1/2\eta_2^2$, $P_{\mathrm{rot}}=\eta_3$, and $\Gamma = 2/\eta_4^2$.}. The $\Gamma$ parameter changes the amplitude of the $\sin^2$ term, and controls the harmonic structure of the resulting GP model \citep[see e.g., discussion in][]{Angus2018}. We follow \cite{haywood2018} and place an informative prior on the $\Gamma$ parameter; \cite{haywood2018} place an informative Gaussian prior of $\eta_4 = 0.5 \pm 0.05$, which corresponds to $\Gamma = 8 \pm 1.9$ (in their notation $\Gamma = 2/\eta_4^2$). In doing so enforces the RV curve to have up to two or three maxima and two or three minima per rotation, as is typical of stellar light curves and RV curves \citep{haywood2018}.

Second, for a computationally efficient analysis of the \textit{K2} data, we elect to use the quasi-periodic kernel available in \texttt{celerite} \citep{Foreman-Mackey2017}, where the kernel function is given with the following form,
\begin{equation}
k(x_l,x_m) = \frac{B}{2 + C} e^{-\tau/L} \left[ \cos \left( \frac{2 \pi \tau}{P_\mathrm{GP}} \right) + (1 + C) \right],
\label{eq:kernelperiodic}
\end{equation}
where $\tau = |x_l - x_m|$, and where $B$, $C$, $L$, and $P_{\mathrm{rot}}$ are the hyperparameters of the kernel. $B$ and $C$ tune the weight of the exponential decay component of the kernel with a decay constant of $L$ (in days), and $P_{\mathrm{GP}}$ corresponds to the periodicity of the quasi-periodic oscillations which we interpret as the stellar rotation period. Although not exactly of the same form as the kernel in Equation \ref{eq:kernelgeorge}, we selected this kernel as it shares similar quasi-periodic properties as the kernel in Equation \ref{eq:kernelgeorge}, but is orders of magnitude faster to evaluate on the large number of \textit{K2} datapoints \citep[see discussion in][]{Foreman-Mackey2017}.Given the high precision and the clear photometric modulation seen in the \textit{K2} data, we share the $P_{\mathrm{GP}}$ parameter between the quasi-periodic \textit{K2} GP kernel and our quasi-periodic RV kernel, to allow the high precision \textit{K2} photometry to accurately constrain the $P_{\mathrm{GP}}$ parameter. We did not share any other GP parameters between the \textit{K2} and the RV kernels, given the different forms of the kernels used. Table \ref{tab:priors1} further lists all of the priors used and which parameters are linked between different GP kernels.

Lastly, the ground-based photometric transit observations are not long enough to be be measurably impacted by the starspot modulation seen in the \textit{K2} light curve. Instead, the characteristic timescales of the observed correlated noise is much shorter or $<1 \unit{hour}$, originating from observational and/or instrumental systematics. Therefore, for the ground-based observations, we adopt the Approximate Matern-3/2 kernel multiplied by an exponential kernel which has covariance properties that are better matched to these timescales \citep[see e.g.,][]{pepper2017,Espinoza2019}. As implemented in \texttt{juliet}, this kernel has the following form (see also \cite{Foreman-Mackey2017}),
\begin{equation}
k(x_l,x_m) = \sigma_{\mathrm{GP}}^2 e^{-\tau/L} \left[ (1 + 1/\epsilon) e^{-1(1-\epsilon)s} + (1 - 1/\epsilon) e^{-1(1+\epsilon)s} \right],
\label{eq:kernelmat32}
\end{equation}
where $s = \sqrt{3} \tau / \rho$, and $\tau = |x_l - x_m|$, with hyperparameters $\sigma_{\mathrm{GP}}$ (photometric amplitude in $\mathrm{ppm}$), $L$ (length scale of the exponential component in days), and $\rho$ (length scale of the Matern-3/2 kernel in days), and with $\epsilon = 0.01$, where we note that as $\epsilon$ approaches 0 the factor inside the brackets converges to a Matern 3/2 kernel \citep{Foreman-Mackey2017,Espinoza2019}. To allow for sufficient flexibility in modeling out different systematics seen in the different observing setups in the ground-based light curves, we assigned each ground-based transit an independent Approximate Matern 3-2 kernel with independent hyperparameters ($\sigma_{\mathrm{GP}}$, $L$, and $\rho$). As an additional test, we also experimented using a pure exponential kernel to model the correlations seen in the ground-based data (with a timescale $L$ and $\sigma$ amplitude parameters) as implemented in \texttt{juliet}. Although both kernels yielded consistent planet parameters within the $68.3\%$ credible intervals of the posteriors ($\sim$$1\sigma$ uncertainties for a Gaussian distribution) in our experiments, the pure exponential kernel tended to favor small decay values that visually over-fit the noise structures in the data. The additional capability of the composite kernel in Equation \ref{eq:kernelmat32} to account for both lower and higher frequency correlated noise in the ground-based light curves led to less over-fitting of the noise \citep[see e.g., additional discussions in][]{pepper2017,Espinoza2019}, and we thus favor this kernel in our analysis.

To investigate the evidence for eccentricity in the system, and to study the impact that our GP RV model has on the derived orbital parameters, we ran three different models. First, we performed a fit assuming a circular orbit ($e=0$) (Model RV1 in Table \ref{tab:priors1}) with a simultaneous GP fit for the transits and radial velocity data. Second, we performed a fit letting the eccentricity $e$ and argument of periastron $\omega$ float (Model RV2 in Table \ref{tab:priors1}), while also performing a simultaneous GP fit like in Model RV1. Third, we let $e$ and $\omega$ float, but removed the GP model on the radial velocity data (keeping the GP model for the photometry; Model RV3 in Table \ref{tab:priors1}). The priors for these three models are summarized in Table \ref{tab:priors1}. In total, we sampled 70, 72 and 69 parameters for the three fits, where we obtained 64077, 54258, and 52522 posterior samples for the three models, respectively.

In all three models we place a Gaussian prior on the stellar density of $\rho = 14.7 \pm 1.5 \unit{g\:cm^{-3}}$ estimated from the mass and radius of the star in Table \ref{tab:stellarparam}. In doing so, allows us to place an important constraint on the eccentricity and argument of periastron derived from the transit from the photo-eccentric effect \citep{dawson2012}. \cite{mann2016} and \cite{thao2020} also report evidence of non-zero eccentricity in the system from their analyses of $e = 0.27_{-0.21}^{+0.16}$, and $e = 0.27_{-0.06}^{+0.16}$, respectively. Our data also favors an eccentric solution. In performing a circular fit placing no constraint on the stellar density, the resulting best-fit fit yielded a stellar density of $\rho = 40 \pm 4 \unit{g\:cm^{-3}}$, significantly larger than the expected stellar density of $\rho = 14.7 \pm 1.5 \unit{g\:cm^{-3}}$. Therefore, to ensure an $a/R_*$ value that results in a stellar density consistent with our estimated stellar density derived from the stars mass and radius, we placed an informative Gaussian prior on the stellar density in all fits reported in Table \ref{tab:planetparams}.

To estimate the expected photometric amplitude from ellipsoidal variations caused by tidal interactions between the planet and the host star, we used Equation 7 in \cite{shporer2017}, which gives the expected ellipsoidal variation amplitude in parts-per-million as a function of the stellar mass, planetary mass, the orbital distance of the planet, and the limb darkening and gravity darkening parameters of the host star. Using our parameter constraints, nominal vales for the limb-darkening and gravity darkening parameters, and assuming a circular orbit with an $a/R_*=21$, we obtain an expected ellipsoidal variation amplitude of 0.03ppm. Even if we assume a circular orbit at K2-25b's periastron distance of $a/R_*$$\sim11.5$, we obtain a small expected ellipsoidal variation amplitude of 0.2ppm. Both values are substantially below the photometric precision of our datasets. Given the low amplitude of the signal, we did not attempt to fit any ellipsoidal variations as part of our transit model.

\begin{deluxetable*}{llcccc}
\tablecaption{Median values and 68\% credible intervals for the three joint-fit models considered. We adopt the values for Model RV2 as it is statistically favored over the other two models.\label{tab:planetparams}}
\tablehead{\colhead{~~~Parameter}&                            \colhead{Description}  &      \colhead{Model RV1}                &        \colhead{\textbf{Model RV2 (adopted)}}    & \colhead{Model RV3} \\
           \colhead{}           &                                 \colhead{}         &     \colhead{($e=0$, with GP for RVs)}  &    \colhead{($e\neq0$, with GP for RVs)}& \colhead{($e\neq0$, no GP for RVs)}}
\startdata
\multicolumn{5}{l}{\hspace{-0.3cm} Model Evidence:}           \\
                        $\ln Z$ &                                 Model log evidence &                         $25828.7\pm9.9$ &                         $25875.4\pm4.1$ &                         $25869.9\pm4.2$ \\ 
\multicolumn{5}{l}{\hspace{-0.3cm} Planet Parameters:}           \\
 $T_{C}$ $(\mathrm{BJD_{TDB}})$ &                                   Transit Midpoint &   $2458515.64215_{-0.00008}^{+0.00008}$ &   $2458515.64206_{-0.00009}^{+0.00010}$ &  $2458515.642134_{-0.00008}^{+0.00008}$ \\
                            $P$ &                              Orbital period (days) &  $3.48456408_{-0.0000005}^{+0.0000005}$ &  $3.48456408_{-0.0000005}^{+0.0000006}$ &  $3.48456407_{-0.0000006}^{+0.0000005}$ \\
                    $(R_p/R_*)$ &                                       Radius ratio &            $0.1146_{-0.0011}^{+0.0010}$ &            $0.1075_{-0.0018}^{+0.0018}$ &             $0.108_{-0.0019}^{+0.0018}$ \\
                        $R_{p}$ &                         Planet radius ($R_\oplus$) &                  $3.66_{-0.12}^{+0.12}$ &                  $3.44_{-0.12}^{+0.12}$ &                  $3.45_{-0.12}^{+0.12}$ \\
                        $R_{p}$ &                              Planet radius ($R_J$) &               $0.327_{-0.011}^{+0.011}$ &               $0.306_{-0.011}^{+0.011}$ &               $0.308_{-0.011}^{+0.011}$ \\
                   $\delta_{p}$ &                                      Transit depth &         $0.01314_{-0.00025}^{+0.00024}$ &         $0.01155_{-0.00039}^{+0.00038}$ &         $0.01167_{-0.00041}^{+0.00040}$ \\
                        $a/R_*$ &                             Scaled semi-major axis &                 $24.47_{-0.33}^{+0.33}$ &                 $21.09_{-0.59}^{+0.57}$ &                 $21.29_{-0.64}^{+0.55}$ \\
                            $a$ &           Semi-major axis (from $a/R_*$ and $R_*$) &            $0.0334_{-0.0011}^{+0.0012}$ &            $0.0287_{-0.0012}^{+0.0012}$ &             $0.029_{-0.0012}^{+0.0012}$ \\
                            $i$ &                     Transit inclination ($^\circ$) &              $88.068_{-0.046}^{+0.046}$ &                 $87.16_{-0.21}^{+0.18}$ &                 $87.24_{-0.23}^{+0.18}$ \\
                            $b$ &                                   Impact parameter &            $0.8252_{-0.0096}^{+0.0092}$ &               $0.628_{-0.037}^{+0.032}$ &               $0.619_{-0.043}^{+0.037}$ \\
                            $e$ &                                       Eccentricity &                     $0.0_{-0.0}^{+0.0}$ &               $0.428_{-0.049}^{+0.050}$ &               $0.409_{-0.039}^{+0.041}$ \\
                       $\omega$ &                  Argument of periastron ($^\circ$) &                    $90.0_{-0.0}^{+0.0}$ &                 $120.0_{-14.0}^{+12.0}$ &                 $106.0_{-16.0}^{+13.0}$ \\
              $T_{\mathrm{eq}}$ &                 Equilibrium temp. (assuming $a=0$) &                   $458.3_{-8.7}^{+8.8}$ &                 $494.0_{-11.0}^{+11.0}$ &                 $492.0_{-11.0}^{+11.0}$ \\
              $T_{\mathrm{eq}}$ &               Equilibrium temp. (assuming $a=0.3$) &                   $320.8_{-6.1}^{+6.2}$ &                   $345.7_{-7.8}^{+8.0}$ &                   $344.3_{-7.8}^{+8.0}$ \\
                            $S$ &                     Insolation Flux ($S_{\oplus}$) &                  $7.35_{-0.55}^{+0.58}$ &                  $9.91_{-0.86}^{+0.95}$ &                  $9.75_{-0.85}^{+0.94}$ \\
                       $T_{14}$ &                            Transit duration (days) &         $0.03398_{-0.00024}^{+0.00027}$ &         $0.03182_{-0.00037}^{+0.00036}$ &         $0.03162_{-0.00035}^{+0.00037}$ \\
                       $T_{23}$ &                            Transit duration (days) &         $0.01454_{-0.00090}^{+0.00091}$ &         $0.02212_{-0.00065}^{+0.00063}$ &         $0.02208_{-0.00071}^{+0.00068}$ \\
                         $\tau$ &                     Ingress/egress duration (days) &         $0.00972_{-0.00042}^{+0.00042}$ &         $0.00485_{-0.00039}^{+0.00040}$ &         $0.00477_{-0.00042}^{+0.00046}$ \\
                            $K$ &                            RV semi-amplitude (m/s) &                    $24.7_{-6.8}^{+7.2}$ &                    $27.9_{-6.0}^{+6.5}$ &                    $32.2_{-9.4}^{+9.7}$ \\
                          $m_p$ &                           Planet mass ($M_\oplus$) &                    $24.0_{-6.6}^{+7.1}$ &                    $24.5_{-5.2}^{+5.7}$ &                    $28.5_{-8.3}^{+8.5}$ \\
        $\sigma_{\mathrm{HPF}}$ &                                HPF RV jitter (m/s) &                    $13.9_{-5.3}^{+6.9}$ &                     $1.8_{-1.4}^{+5.1}$ &                    $42.9_{-6.5}^{+7.2}$ \\
                       $\gamma$ &                                HPF RV offset (m/s) &                  $-5.0_{-12.0}^{+13.0}$ &                   $4.0_{-16.0}^{+16.0}$ &                    $-4.0_{-7.5}^{+7.5}$ \\
\multicolumn{5}{l}{\hspace{-0.3cm} Derived Stellar Parameters:}           \\
                         $\rho$ &              Stellar density ($\unit{g\:cm^{-3}}$) &                 $22.83_{-0.91}^{+0.92}$ &                    $14.6_{-1.2}^{+1.2}$ &                    $15.0_{-1.3}^{+1.2}$ \\
\multicolumn{5}{l}{\hspace{-0.3cm} GP Hyperparameters for RVs:}           \\
         $\sigma_{\mathrm{GP}}$ &                              GP RV amplitude (m/s) &                    $34.8_{-6.6}^{+8.6}$ &                   $41.0_{-7.9}^{+11.0}$ &                    - \\
                       $\alpha$ &           GP inverse timescale ($\unit{day^{-2}}$) &           $13_{-10}^{+90}\times 10^{-6}$ &           $15_{-1}^{+70}\times10^{-8}$ &                    - \\
                       $\Gamma$ &                  GP Frequency Structure parameter  &                     $8.0_{-1.1}^{+1.1}$ &                   $6.75_{-1.2}^{+0.93}$ &                    - \\
              $P_{\mathrm{GP}}$ &             GP Kernel Periodicity parameter (days) &         $1.88219_{-0.00032}^{+0.00034}$ &         $1.88203_{-0.00031}^{+0.00032}$ &                    - \\
\enddata
\end{deluxetable*}

\begin{figure*}[t!]
\begin{center}
\includegraphics[width=\textwidth]{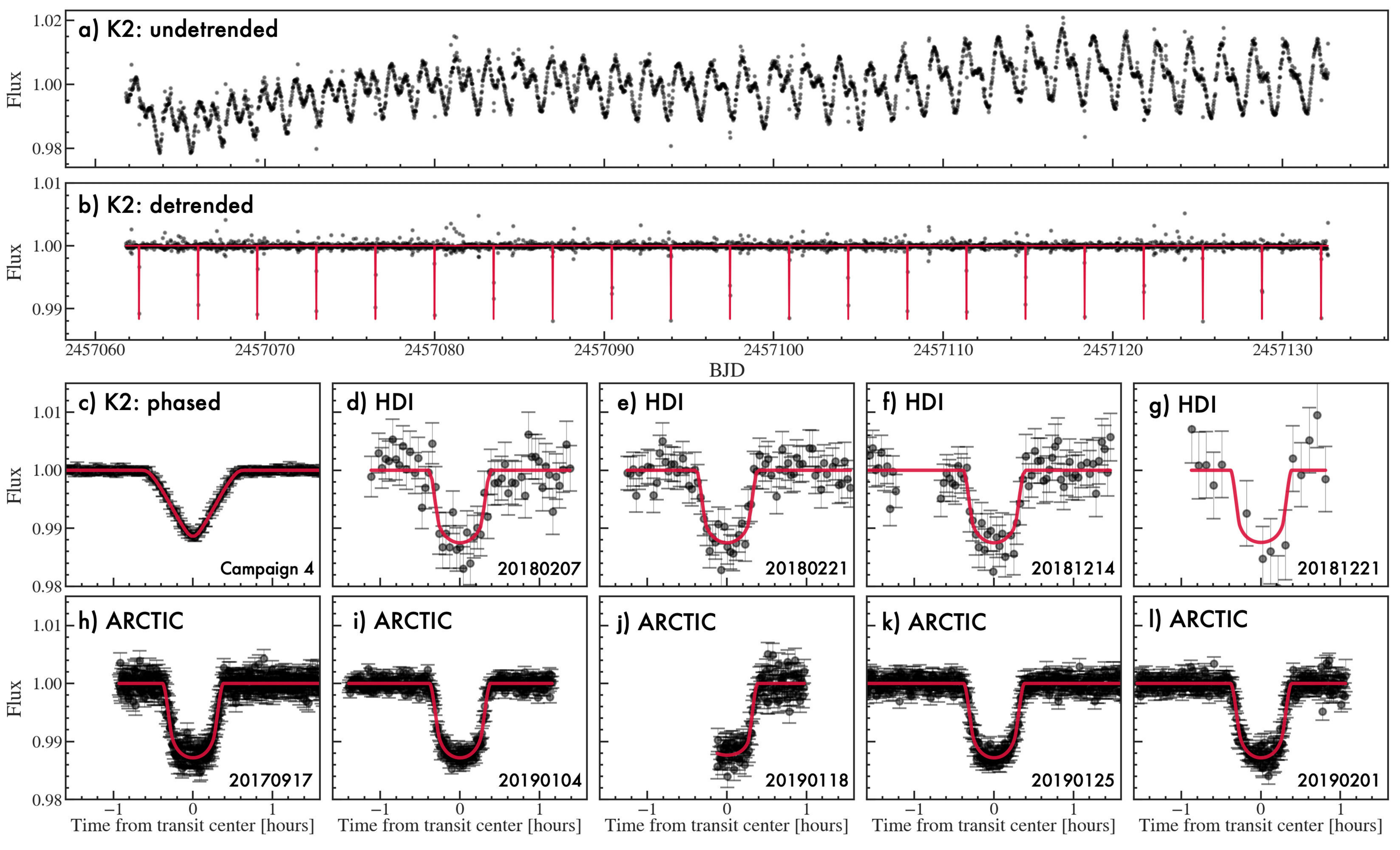}
\vspace{-0.5cm}
\end{center}
\caption{a) \textit{K2} light curve of K2-25 corrected for instrument systematics using the \texttt{Everest} pipeline, showing clear starspot induced rotational modulations. b) Detrended and flattened \textit{K2} light curve shown in black points using the best-fit Gaussian Process model from Model RV2 in Table \ref{tab:planetparams}, along with the best-fit transit model from \texttt{Juliet} shown in red. c) Phased \textit{K2} light curve (black) along with the best-fit transit model (red). d)-g) Ground-based transits as observed with HDI showing the best-fit transit model from \texttt{Juliet} in red. We note that the HDI transit on 20181221 (g) was observed without the diffuser during sub-optimal observing conditions (Moon illumination 97\%). Although the light curve was modelled unbinned, it is shown here binned to a cadence of $\sim$6 minutes for clarity. h-l) Ground-based transits observed with the 3.5m ARC Telescope at Apache Point. All of the ARCTIC observations were observed using a diffuser, except the transit on the night of 20190118 UT (j).}
\label{fig:transits}
\end{figure*}

\subsection{Derived Parameters}
Figure \ref{fig:transits} shows the \textit{K2} transits along with the 9 ground-based transits observed and analyzed in this work. The transits on nights of December 21, 2018 and January 4, 2019 were performed simultaneously with the RM effect observations further discussed in Section \ref{sec:rm}. The shorter cadence of the ground-based observations compared to the 30 minute cadence \textit{K2} observations allows us to resolve the transit shape better, resulting in more precise planet parameters. Table \ref{tab:planetparams} compares the best-fit parameters from the three models considered in this work.

\begin{figure*}[t]
\begin{center}
\includegraphics[width=1.0\textwidth]{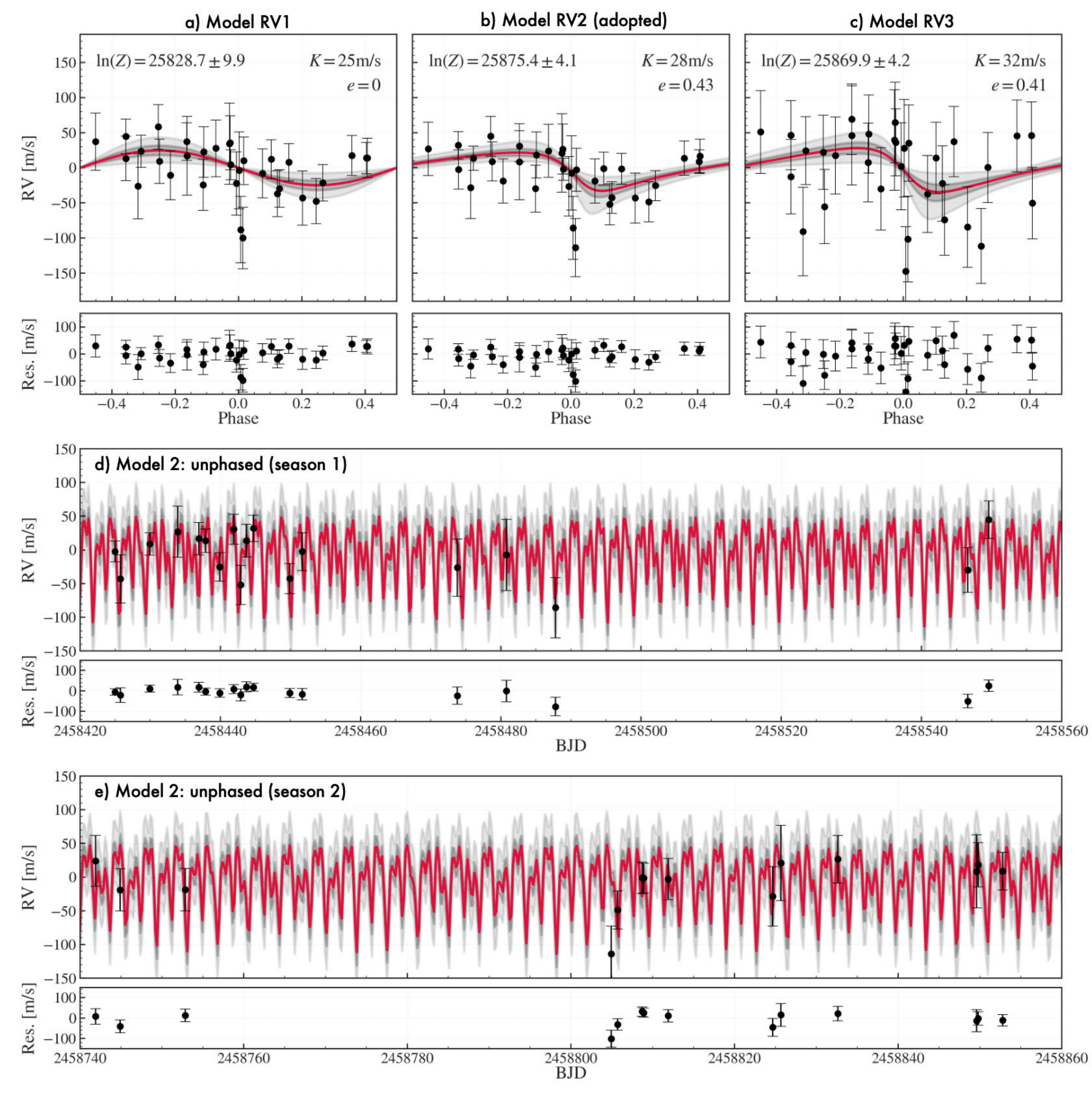}
\caption{Out-of-transit RVs of K2-25 from HPF comparing the resulting RV models from a) Model RV1, b) Model RV2, and c) Model RV3, with associated residuals shown in the corresponding lower panels. The 50th percentile best-fit models are shown in red. The grey shaded regions show the $1\sigma$ (grey) and $3 \sigma$ (lighter grey) estimated model confidence bands. The log-evidence for the three models are $\ln Z = 25828.7 \pm 9.9$, $\ln Z = 25875.4 \pm 4.1$, and $\ln Z = 25869.9 \pm 4.2$, respectively, showing a significant preference for the eccentric models (Models RV2 and RV3). We adopt Model RV2 due to the statistical preference over Model RV3, with $\Delta \ln Z \sim 5$ over Model RV3. d-e) The non-phased RVs as a function of time for two observing Seasons (approximately a year apart). The red curve shows the best-fit GP and Keplerian model. The grey shaded regions show the $1\sigma$ (grey) and $3 \sigma$ (lighter grey) estimated model confidence bands.}
\label{fig:rvs}
\end{center}
\end{figure*}

Figure \ref{fig:rvs} compares the resulting phased RV plots for the three different models considered. The derived best-fit RV semi-amplitudes for the three models are $K_{\mathrm{RV1}} = 24.7_{-6.8}^{+7.2} \unit{m/s}$, $K_{\mathrm{RV2}} = 27.9_{-6.0}^{+6.5} \unit{m/s}$, and $K_{\mathrm{RV3}} = 32.2_{-9.4}^{+9.7} \unit{m/s}$, respectively. This results in mass estimates of $M_{\mathrm{RV1}} = 24.0_{-6.6}^{+7.1} \unit{M_\oplus}$, $M_{\mathrm{RV2}} = 24.5_{-5.2}^{+5.7} \unit{M_\oplus}$, and $M_{\mathrm{RV3}} = 28.5_{-8.3}^{+8.5} \unit{M_\oplus}$ for the three models, respectively. We see that all three models result in consistent mass estimates. We note that the GP RV amplitudes for Model RV1 and RV2 are similar with $\sigma_{\mathrm{GP,RV1}} = 35_{-7}^{+9} \unit{m/s}$, and $\sigma_{\mathrm{GP,RV2}} = 41_{-8}^{+11} \unit{m/s}$. As expected, we see that for Model RV3---which does not employ a GP---the white noise error term significantly increases to compensate for the additional correlated-noise, with $\sigma_{\mathrm{w,RV1}} = 13.9_{-5.3}^{+6.9} \unit{m/s}$, $\sigma_{\mathrm{w,RV2}} = 1.8_{-1.4}^{+5.1} \unit{m/s}$, and $\sigma_{\mathrm{w,RV3}} = 42.9_{-6.5}^{+7.2} \unit{m/s}$.

To check which model between models RV1, RV2, and RV3, is statistically favored, we use the log-evidence values calculated by the \texttt{dynesty} dynamic nested sampler. As the estimated errors of the log-evidences can be underestimated by nested sampling algorithms \citep[see e.g.,][]{nelson2018,Espinoza2019}, to get an accurate estimate of the distribution in the log-evidence values for each model, we ran each joint-fit model RV1, RV2, RV3, six separate times. In doing so, we obtain log evidence values of $\ln Z = 25828.7 \pm 9.9$, $\ln Z = 25875.4 \pm 4.1$, and $\ln Z = 25869.9 \pm 4.2$, respectively, where the value reported is the mean of the 6 runs for each model and the uncertainty estimate is the standard deviation of the 6 runs. We note that the scatter in the log-evidence values ($\sim$10 for Model RV1, and $\sim$$4$ for Models RV2 and RV3) is substantially larger than the internal uncertainty estimate of $\sim$$0.5$ reported from the \texttt{dynesty} sampler for each individual run. Even so, within each group of 6 runs for Models RV1, RV2, and RV3, the resulting output posteriors are fully consistent. Between the three models, we see that the two eccentric Models RV2 and RV3 are favored over the circular Model RV1. Between Models RV2 and RV3, we see a statistical preference ($\Delta \ln Z \sim 5$) in favor of Model RV2. From Table \ref{tab:planetparams}, we see that the derived planet parameters, including the semi-amplitude, mass, and eccentricity for Models RV2 and RV3 are fully consistent. Given the statistical preference, and the fact we know that K2-25 is a young and active star with a well characterized stellar rotation period that is explicitly modelled in model RV2, we elect to adopt the values from Model RV2. We compared the distribution of the best-fit residuals of the HPF RVs from model RV2 to a normal distribution using the Kolmogorov-Smirnov test, from which we obtain a $p$-value of $6.5\times10^{-8}$. This suggests that the distribution of the RV residuals is indistinguishable from a normal distribution, and that our model (1 planet Keplerian along with a quasi-periodic GP) can accurately model the observed RVs.

\subsection{Stellar Activity from HPF Spectra}
\begin{figure}
\begin{center}
\includegraphics[width=1.0\columnwidth]{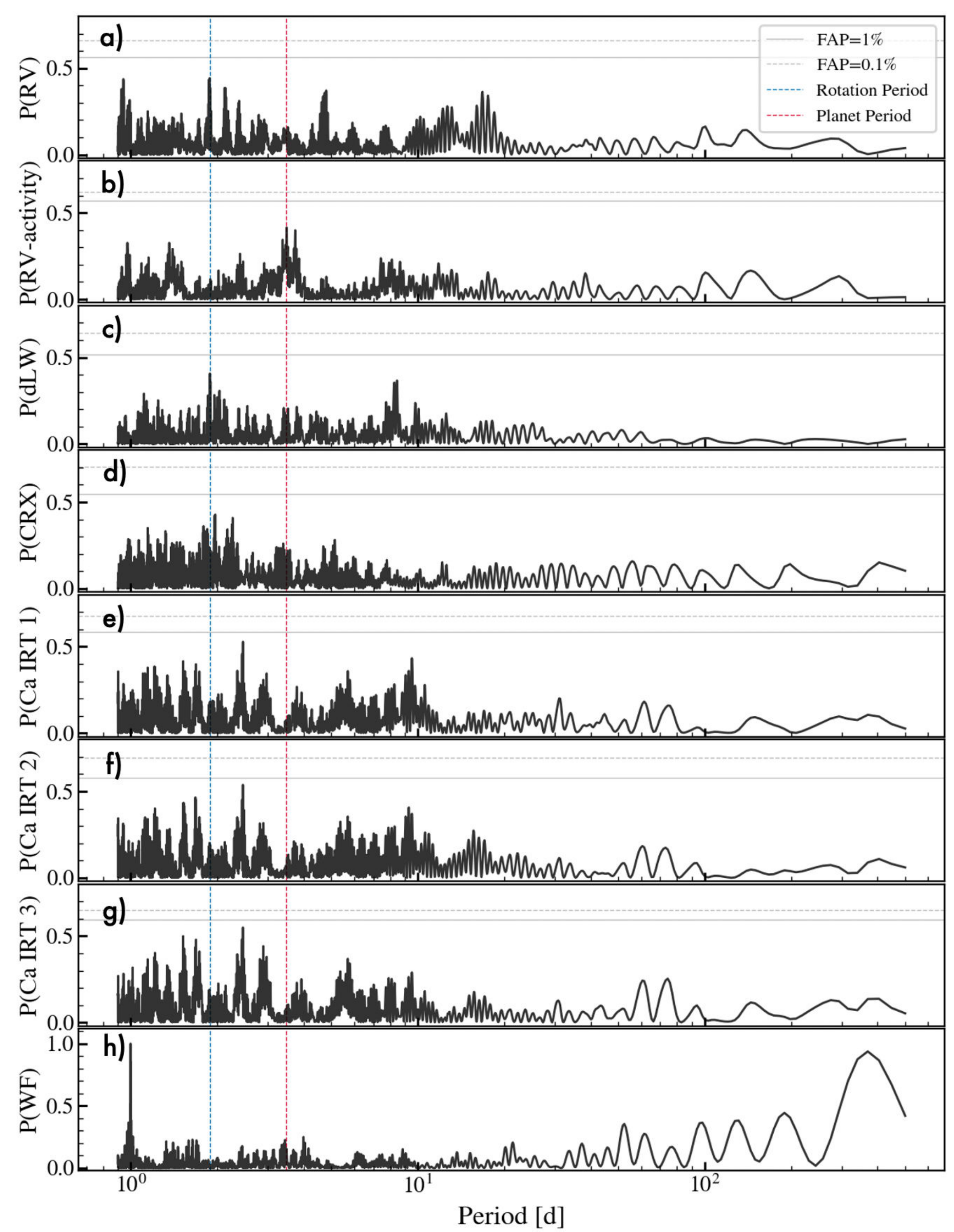}
\caption{Lomb-Scargle periodograms of the HPF RVs along with different activity indicators. The stellar rotation period ($P=1.88 \unit{days}$) and the planet period ($P=3.48 \unit{days}$) are highlighted with the dashed blue and red lines, respectively. False alarm probabilities of 1\% and 0.1\% calculated using a bootstrap method are denoted with the grey solid and grey dashed lines, respectively. a) the HPF out-of-transit RVs used for the mass measurement of K2-25b. b) same as a) but after removing the best-fit GP model from Model RV2. This shows that the known planet peak becomes more significant. c) Differential line width (dLW) activity indicator, showing a clear peak at the known rotation period. d) Chromatic Index activity indicator. e-g) Ca II IRT indices for the three Ca II IRT lines. h) The window function of the HPF RVs, showing a clear sampling peak at 1 day. The power in a-g) is normalized using the formalism in \cite{zechmeister2009}, and h) is normalized so that the highest peak is unity.}
\label{fig:periodograms}
\end{center}
\end{figure}

To further study the activity of the star, we measured a number of stellar activity indicators from the HPF spectra. Figure \ref{fig:periodograms} shows the Generalized Lomb-Scargle (LS) periodograms of the out-of-transit RVs we used for our mass measurement, along with an array of different activity indicators measured from the HPF spectra, including: the Differential Line Width (dLW), the Chromatic Index (CRX), and line indices of the three Calcium II Infrared Triplet (Ca II IRT) lines. To calculate the LS periodograms, we used the periodogram functions in the \texttt{astropy.timeseries} package, and the false alarm probabilities\footnote{Although the False Alarm Probability is a commonly used in periodogram analysis in radial velocity data it has known limitations \citep[see e.g., discussion in][]{fischer2016}.} were calculated using the \texttt{bootstrap} method implemented in this same package. Additionally, in Figure \ref{fig:periodograms}, we show the Window Function (WF) of our RV observations. All of the periodograms in Figure \ref{fig:periodograms} are normalized using the formalism in \cite{zechmeister2009}, except the window function is normalized such that the highest peak has a power of 1. Table \ref{tab:rvsout} in the appendix lists the values of the RVs and the activity indicators used in this work.

To measure these indicators, we follow the definition and procedures in the SERVAL pipeline \citep{zechmeister2018}. To measure the Ca II IRT indicators, we measure the mean flux in a 30km/s wide region centered on the three Ca II IRT line centers and we use two offset bands (to the right and to the left of the line center, 100km/s wide) as reference regions to measure line indices as defined in Equation 27 in \citep{zechmeister2018},
\begin{equation}
I = \frac{\langle f_0 \rangle}{0.5(\langle f_1\rangle + \langle f_2\rangle)},
\label{eq:index}
\end{equation}
where $\langle f_0 \rangle$ is the mean flux around the line center, and $\langle f_1 \rangle$ and $\langle f_2 \rangle$ are the mean fluxes around the reference regions. The exact locations of the line center regions and the offset regions we used are given in Table \ref{tab:cairt} in the Appendix.

From Figure \ref{fig:periodograms}a, we see a clear peak in the RV periodogram at the known rotation period, indicative of rotationally modulated RV variations e.g., due to starspots. This interpretation is further strengthened by the fact that we see a clear peak in the periodogram of the dLW indicator at the known rotation period (Figure \ref{fig:periodograms}c). In Figure \ref{fig:periodograms}b, we show the periodogram of the RVs after subtracting the best-fit GP activity model, demonstrating that after removing the GP model, the peak at the stellar rotation is significantly suppressed and the peak at the planet period increases in significance. The CRX indicator does not show any clear evidence of periodic variations in the chromaticity of the RVs at either the stellar rotation period or the planet period. Interestingly, in Figures \ref{fig:periodograms}e, f, and g, we do not see clear peaks in the Ca II IRT activity indicators at the rotation period, but rather, we see a clear peak in all three indices at 2.46 days. We speculate that this could indicate that the active chromospheric regions the Ca II IRT lines trace, could have a different characteristic evolution timescale than the rotation period of the star.

\subsection{RV Injection and Recovery Tests}
\label{sec:injection}
To test that our RV precision and the RV sampling is sufficient to accurately constrain the Keplerian parameters of K2-25b, we performed two series of injection and recovery tests, broadly following the methodology in \cite{klein2020}. For both series of tests, we injected a signal with known Keplerian parameters ($P$, $T_C$, $K$, $\omega$, $e$), and Gaussian process hyper-parameters ($P_{\mathrm{GP}}$, $\sigma_{\mathrm{GP}}$, $\alpha_{\mathrm{GP}}$, $\Gamma_{\mathrm{GP}}$), along with a white noise parameter ($\sigma_{\mathrm{w}}$). The values of the parameters for the two tests, along with the priors used for the recovery, are given in Table \ref{tab:injection} in the Appendix. We placed informative Gaussian priors on the orbital period ($P$), transit center ($T_C$), and the GP period ($P_{\mathrm{GP}}$), and the GP frequency structure parameter ($\Gamma_{\mathrm{GP}}$), and we placed broad uninformative priors on the other parameters (same priors as used for fit RV2). For the first series of tests (Test I), we set the GP amplitude $\sigma_{\mathrm{GP}} = 42 \unit{m/s}$, and other parameters to values similar to the nominal 50th quantile values from our adopted RV2 fit in Table \ref{tab:planetparams} (see exact values injected in Table \ref{tab:injection} in the Appendix). For the second series of tests (Test II), we increased the injected GP amplitude to its 95th percentile value from our adopted RV2 model in Table \ref{tab:planetparams}, corresponding to $\sigma_{\mathrm{GP}} = 64.6 \unit{m/s}$, to check if the injected Keplerian parameters---in particular $K$ and $e$---could be reliably recovered if the level of the correlated noise is higher. For both tests, we repeated the injection and recovery 200 times, and we then inspect the resulting posteriors calculated using \texttt{juliet} from each individual run.

Figure \ref{fig:injection} in the Appendix shows the distribution of mean values from Test I for a few select parameters of interest: $K$, $\omega$, $e$, $\sigma_{\mathrm{GP}}$, $\alpha_{\mathrm{GP}}$, and $\sigma_{\mathrm{w}}$. We see that for all parameters, the distribution of the recovered values is fully consistent with the known injected value. Further, Figure \ref{fig:posteriors} in the Appendix compares the distribution of all 200 posteriors for $\sigma_{\mathrm{GP}}$, $K$, and $e$, for both series of injection and recovery tests. To compare the distribution of the synthetic residuals to a normal distribution, we used the Kolmogorov-Smirnov test, from which we see that all datasets have a $p$-value $<10^{-5}$ (with most datasets showing a $p$ value $<10^{-7}$), suggesting that the best-fit residuals from the synthetic tests are indistinguishable from a normal distribution. Although $\sigma_{\mathrm{GP}}$ shows broad posteriors and is relatively poorly constrained, in both series of tests, the true values of $K$ and $e$ are consistently recovered. For Test I, which assumes $\sigma_{\mathrm{GP}}=42 \unit{m/s}$, the true value of $K$ is within the 68\% and 95\% credible intervals in 80\% and 99.7\% of the cases, respectively. For Test II, even with the higher injected value of $\sigma_{\mathrm{GP}}=64.6 \unit{m/s}$, the true value of $K$ is reliably recovered within the 68\% and 95\% credible intervals in 75\% and 97\% of the cases, respectively. We conclude that the number and the sampling of the HPF RVs is sufficient to accurately constrain the Keplerian planet parameters.

\section{The RM effect of K2-25}
\label{sec:rm}

\subsection{RM model}
We model the Rossiter-McLaughlin effect using the prescription given in \cite{hirano2010} and \cite{hirano2011}. Specifically, we use Equation 26 from \cite{hirano2011}, which gives the RM velocity anomaly as,
\begin{equation}
\Delta v = - \left( \frac{2 (\beta + \sigma_{\mathrm{RM}})^2}{2\beta^2 + \sigma_{\mathrm{RM}}^2} \right)^{3/2} f v_p \left( 1 - \frac{v_p^2}{(2\beta^2 + \sigma_{\mathrm{RM}}^2) } + \frac{v_p^4}{2 (2\beta^2 + \sigma_{\mathrm{RM}}^2)^2 } \right),
\label{eq:hirano}
\end{equation}
where $\beta$ indicates the best-fit Gaussian dispersion (in $\unit{km/s}$) of the intrinsic line broadening in the absence of stellar rotation, and $\sigma_{\mathrm{RM}}$ indicates the Gaussian line width component arising from stellar rotation (see \citealt{hirano2010} for details). Here we set $\beta$ to the width of the HPF resolution element, i.e., $\beta = 5.45 \pm 0.5 \unit{km/s}$, where the errorbar is to account for any effects of macroturbulence and/or other non-stellar rotation processes that could broaden the line profile. The $f$ parameter denotes the fraction of the star being blocked by the planet during the transit as a function of time (i.e., $f=1-F$, where $F$ is the photometric transit model), and $v_p$ denotes the sub-planet velocity (in \unit{km/s}) i.e., the velocity of the star being blocked by the planet as a function of time during the transit. As discussed in \cite{hirano2011}, $\sigma_{\mathrm{RM}}$ describes the dispersion of a Gaussian approximating the stellar rotational kernel, and here we follow \cite{hirano2010} and assume $\sigma_{\mathrm{RM}} = v \sin i_* / 1.31$. We neglect any differential rotation, as \cite{dmitrienko2017} showed that the differential rotation of K2-25b is small, or $\Delta \Omega = 0.0071 \pm 0.002 \unit{rad/day}$, and thus negligible during the transit.

\begin{table}[b!]
\centering
\caption{Median values and associated 16th and 84th percentile values from our RM fit.}
\begin{tabular}{llc}
\hline\hline
Parameter  &  Description  & Value \\ 
\hline
\multicolumn{3}{l}{\hspace{-0.3cm} Model Likelihood Parameters:}     \\
$\chi^2_\nu$                                            &   Reduced $\chi^2$                      &                 1.04                   \\
DOF                                                     &   Degrees of freedom                    &                 25                     \\
\multicolumn{3}{l}{\hspace{-0.3cm} MCMC Parameters:}      \\
  $\lambda$                                             & Sky-projected obliquity ($^\circ$)      &               $3 \pm 16$               \\
 $v\sin i_*$                                            & Rotational velocity ($\unit{km/s}$)     &               $8.9 \pm 0.6$            \\
 $\gamma_1$                                             & RV offset ($\unit{m/s}$)                &               $-10_{-28}^{+28}$        \\
 $\gamma_2$                                             & RV offset ($\unit{m/s}$)                &               $58_{-26}^{+27}$         \\
 $\gamma_3$                                             & RV offset ($\unit{m/s}$)                &               $-95_{-28}^{+29}$        \\
     $q_1$                                              & Linear Limb darkening parameter         &               $0.47_{-0.33}^{+0.35}$   \\
     $q_2$                                              & Quadratic Limb darkening parameter      &               $0.44_{-0.31}^{+0.36}$   \\
    $\beta$                                             & Intrinsic stellar line width            &               $5.41_{-0.50}^{+0.49}$   \\
\multicolumn{3}{l}{\hspace{-0.3cm} Derived Parameters:} \\
   $\psi$                                               & 3D obliquity ($^\circ$)                 &               $17_{-8}^{+11}$ \\
\hline
\end{tabular}
\label{tab:rmparams}
\end{table}

\begin{figure*}[t!]
\begin{center}
\includegraphics[width=0.95\textwidth]{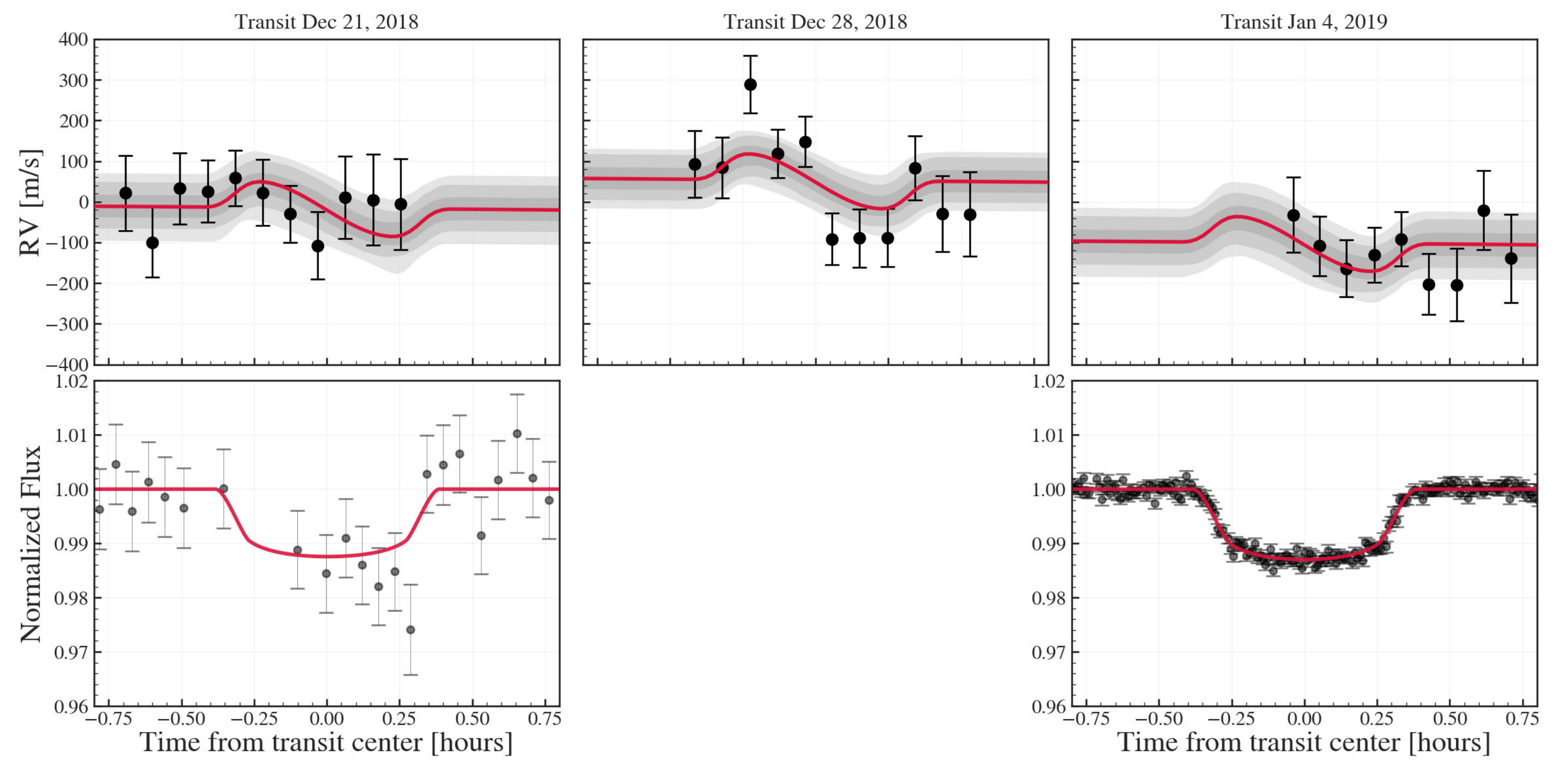}
\caption{Upper panels: Three RM effect transits as observed with HPF (black points), along with the best-fit joint RM effect model shown in red. The RM model includes the bulk RV shift and an independent RV offset parameter $\gamma$ is included for each transit to account for stellar activity and/or instrumental effects. The grey shaded regions show the $1\sigma$, $2\sigma$, and $3\sigma$ confidence regions. Lower panels: Two simultaneous photometric transits from HDI (Dec 21, 2018), and ARCTIC (Jan 4, 2019) shown in black, along with the best-fit transit model (model RV2 in Table \ref{tab:priors1} and \ref{tab:planetparams}). The HDI observations were performed without a diffuser due to the Moon contamination that night. No starspot crossing events are seen in the simultaneous transit observations, which otherwise would complicate the RM-effect analysis. Figure \ref{fig:rmcombined} shows the three RM effect observations phased together and binned to 5 minute bins for further visual inspection.}
\label{fig:rmobs}
\end{center}
\end{figure*}

To model the RM effect, we model all three transits jointly. For the fit, we included the bulk RV motion imposing informative priors on the Keplerian orbital parameters from our best-fit orbital values from Model RV2 in Table \ref{tab:planetparams}. In addition, we placed an informative Gaussian prior on the $v \sin i_*$ using our $v \sin i_*$ value measured from the HPF spectra. To account for possible systematics between the transits (e.g., due to stellar activity and/or instrumental systematics), we added an independent RV offset parameter $\gamma$ for each individual transit. We sampled the limb-darkening parameters using the $q_1$ and $q_2$ parametrization described in \cite{kipping2013}, and fully sample them across the whole valid range from 0 to 1 to minimize any biases on the RM effect anomaly due to limb-darkening effects. To account for the smoothening of the finite exposure times of our RV observations, we supersampled the model 7 times and resampled the model according to the exposure time of $\sim$300s. We calculated the bulk RV model using the \texttt{radvel} python package \citep{fulton2018}. To calculate the transit fraction $f$, we used the \texttt{batman} package \citep{kreidberg2015}, calculating the transit flux and setting $f = 1 - F$, where $F$ is the transit model from \texttt{batman}. Before we started the MCMC sampling, we found the global most probable solution using the \texttt{PyDE} differential evolution optimizer \citep{pyde}. We then initialized 100 MCMC walkers in the vicinity of the global most probable solution using the \texttt{emcee} MCMC affine-invariant MCMC sampling package \citep{dfm2013}. We ran the 100 walkers for 25,000 steps, and after removing the first 2,000 steps as burn-in and thinning the chains by a factor of 100, the Gelman-Rubin statistic of the resulting chains was within $<1\%$ of unity, which we consider well-mixed.

To test the impact of our assumptions about stellar activity on $\lambda$, we performed four additional fits: 
\begin{itemize}
\item A joint fit with a single RV offset parameter $\gamma$.
\item A joint fit with a single RV offset parameter $\gamma$ after subtracting our best-fit GP activity model (from Model RV2 in Table \ref{tab:planetparams}) from the RVs.
\item A joint fit with three independent RV offset parameters $\gamma$ and removing the prior on the semi-amplitude.
\item A joint fit with three independent RV offset parameters $\gamma$ and removing the prior on the $v \sin i_*$.
\end{itemize}

All fits resulted in fully consistent constraints on $\lambda$ suggesting a well-aligned system, although we note that the fit with a single RV offset parameter resulted in slightly higher uncertainty estimate on $\lambda$, or a constraint of $\lambda = 0 \pm 24^\circ$. From the GP activity model constrained from our out-of-transit RVs, we note that the expected RV variation during the three transits observed is slowly varying and significantly smaller ($<5 \unit{m/s}$) than the observed RM amplitude of $\sim$$65 \unit{m/s}$, and thus is effectively modeled out with independent offsets between the transits. In comparing the Baysian Information Criterion (BIC) and Akaike Information Criterion \citep[AIC; ][]{akaike1974}, the models show a statistical preference ($\Delta$AIC$\sim$9) for models that allowed for independent offsets between the transits. We note that using the BIC and AIC in this case is heuristic, as more datapoints would be formally needed for these criteria to be in the asymptotic regime where they are accurate and justified for model comparison. To allow for flexibility to take out potential systematic offsets between the three transits, we report the resulting posterior constraint from the fit assuming three independent RV offset $\gamma$ parameters in Table \ref{tab:rmparams}. We further note that the fit where we allowed $v \sin i_*$ to vary freely resulted in a fully consistent $\lambda$ value and yielded a $v \sin i_* = 11.6 \pm 3 \unit{km/s}$, which is consistent with the as-measured $v \sin i_* = 8.8 \pm 0.6 \unit{km/s}$ from the HPF-spectra at the $1\sigma$ level. 

\subsection{Results}
\begin{figure}[t!]
\begin{center}
\includegraphics[width=0.95\columnwidth]{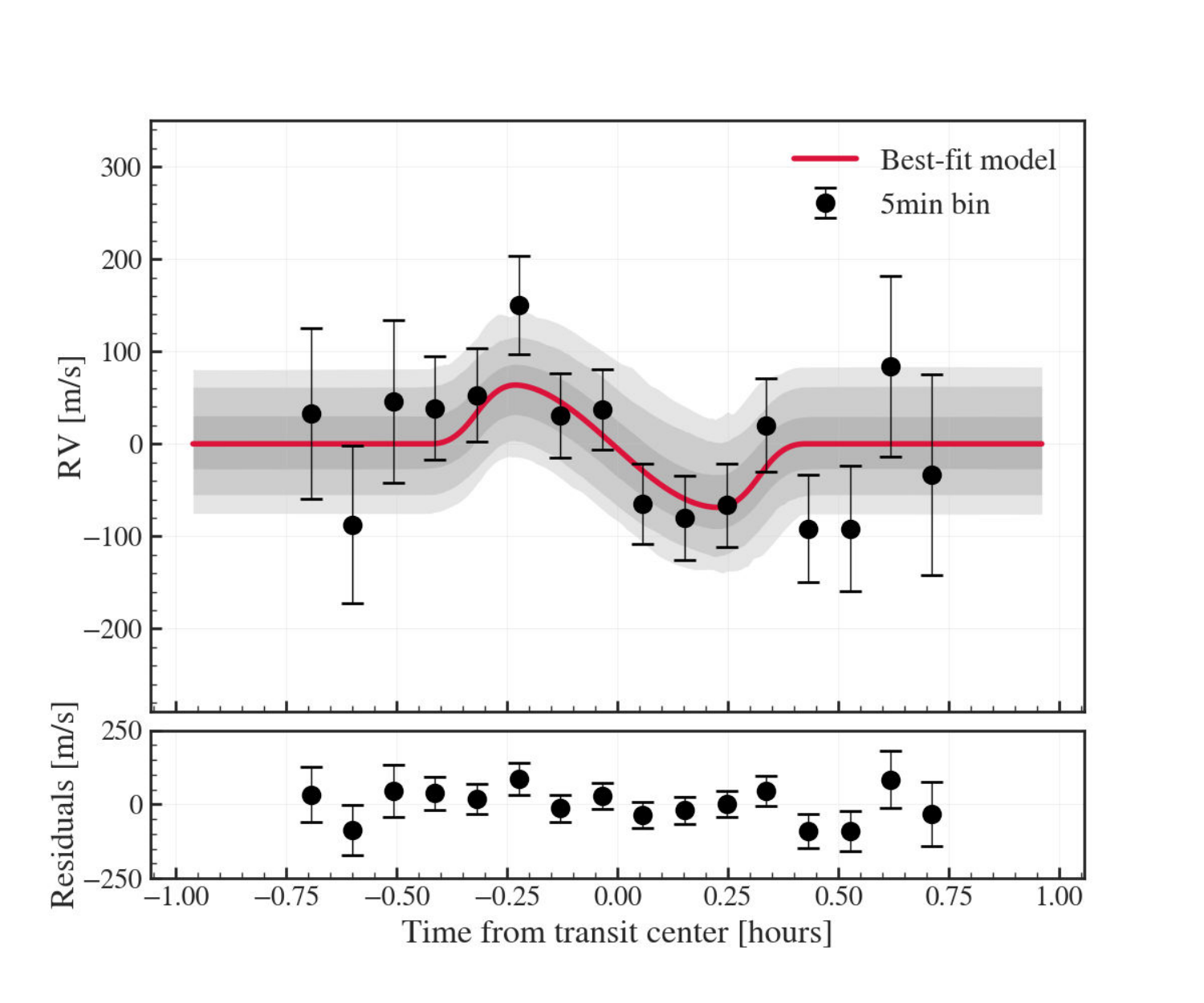}
\vspace{-0.2cm}
\caption{The RM effect of K2-25b after combining the three jointly fitted RM effect observations shown in the upper panels in Figure \ref{fig:rmobs}. The best-fit RM model (red curve) has the bulk RV motion and RV offset parameters removed. The black points show the data binned to a 5min cadence. The grey shaded regions show the 1$\sigma$, 2$\sigma$ and 3$\sigma$ confidence bands.}
\label{fig:rmcombined}
\end{center}
\end{figure}

Figure \ref{fig:rmobs} shows the three RM effect observations using HPF along with our best-fit RM model (RM amplitude $\sim$$65 \unit{m/s}$), along with 1$\sigma$, 2$\sigma$, and 3$\sigma$ shaded regions. Table \ref{tab:rmparams} summarizes the resulting best-fit posterior values, showing that we obtain a sky-projected obliquity constraint of $\lambda = 3 \pm 16^\circ$. To further visualize and compare the observed RVs to the best-fit RM model, in Figure \ref{fig:rmcombined} we show all of the three transits in Figure \ref{fig:rmobs} phased to the transit ephemeris and binned to a 5 minute cadence using a weighted average. The resulting median RV error is $55 \unit{m/s}$ in the 5 minute bins. The RM model is shown with the bulk RV model and RV offset parameters removed for clarity.

In addition to the sky-projected obliquity listed in Table \ref{tab:rmparams}, we calculate the true obliquity angle $\psi$, using the following equation,
\begin{equation}
\cos \psi = \sin i_* \cos \lambda \sin i + \cos i_* \cos i,
\label{eq6:psi}
\end{equation}
where $i_*$ is the stellar inclination, $i$ is the transit inclination, and $\lambda$ is the sky-projected obliquity angle. Using Equation \ref{eq6:psi} above and our $\lambda=3 \pm 16^\circ$ constraint from Table \ref{tab:rmparams}, our $i=87.13\pm0.17^\circ$ from Table \ref{tab:planetparams} (Model RV2), and our $i_* = 90 \pm 12^\circ$ constraint from Table \ref{tab:stellarparam}, we obtain the following constraint on $\psi = 17_{-8}^{+11\:\circ}$. Integrating the resulting posterior, we can say that $\psi < 30^{\circ}$ with 89\% confidence, which is compatible with a well aligned system.

\section{Discussion} 
\label{sec:discussion}
With its known age and characterized 3D orbital architecture, K2-25b is an interesting laboratory to test different formation scenarios. Below, in Section \ref{sec:ephemeris}, we compare our improved ephemerides to the ephemerides of other recent literature on this system. In Section \ref{sec:composition}, we discuss the composition of K2-25b showing that at its observed mass and radius it is consistent with a water-rich world, or a rocky core with a small H/He envelope. In Section \ref{sec:transmission}, using our mass constraint we discuss the possibility for atmospheric characterization through transmission spectroscopy with JWST in the future. We discuss our obliquity constraint in Section \ref{sec:alignment}, which is suggestive of a well-aligned orbit, and place it in context to other measurements in the literature. Finally, given our detailed characterization of the different orbital parameters of the planet allows us to place informative constraints on different formation scenarios, which we discuss in Section \ref{sec:formation}.

\begin{figure}[t!]
\begin{center}
\includegraphics[width=0.9\columnwidth]{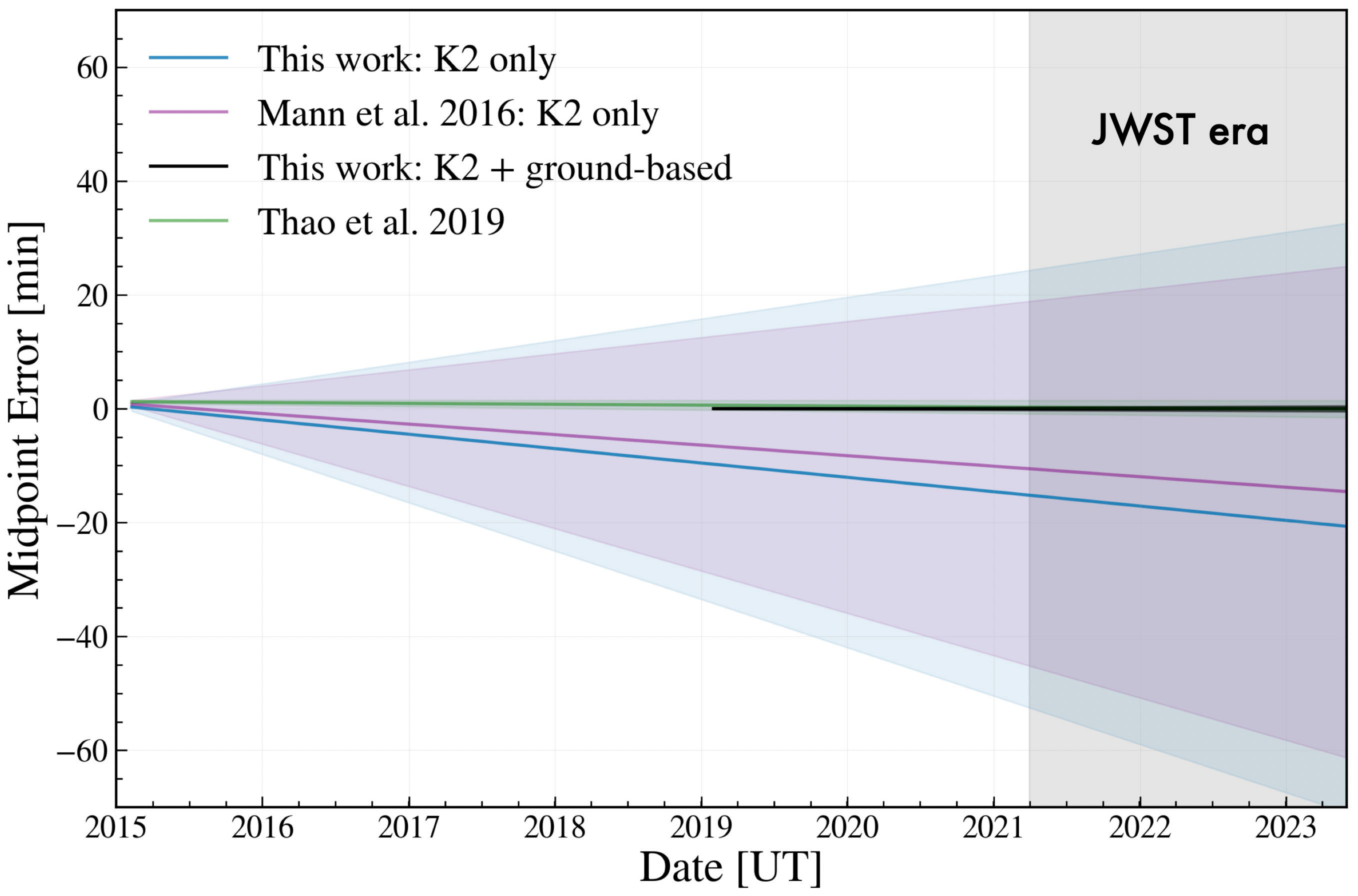}
\caption{Updated transit ephemeris for K2-25b into the JWST era. Our ephemeris derived from the \textit{K2} photometry only (blue shaded region) agrees well with the ephemeris provided in \cite{mann2016} (purple). Our improved ephemeris (Model RV2 in Table \ref{tab:planetparams}) results in an error of $\sim$20sec at the beginning of the JWST era in nominally 2021, and is in excellent agreement with the ephemeris provided in \cite{thao2020} derived from data from \textit{K2}, \textit{Spitzer}, and additional ground-based observations. The shaded regions show the $1\sigma$ error estimates.}
\label{fig:ephemeris}
\end{center}
\end{figure}

\begin{figure*}[t!]
\begin{center}
\includegraphics[width=0.75\textwidth]{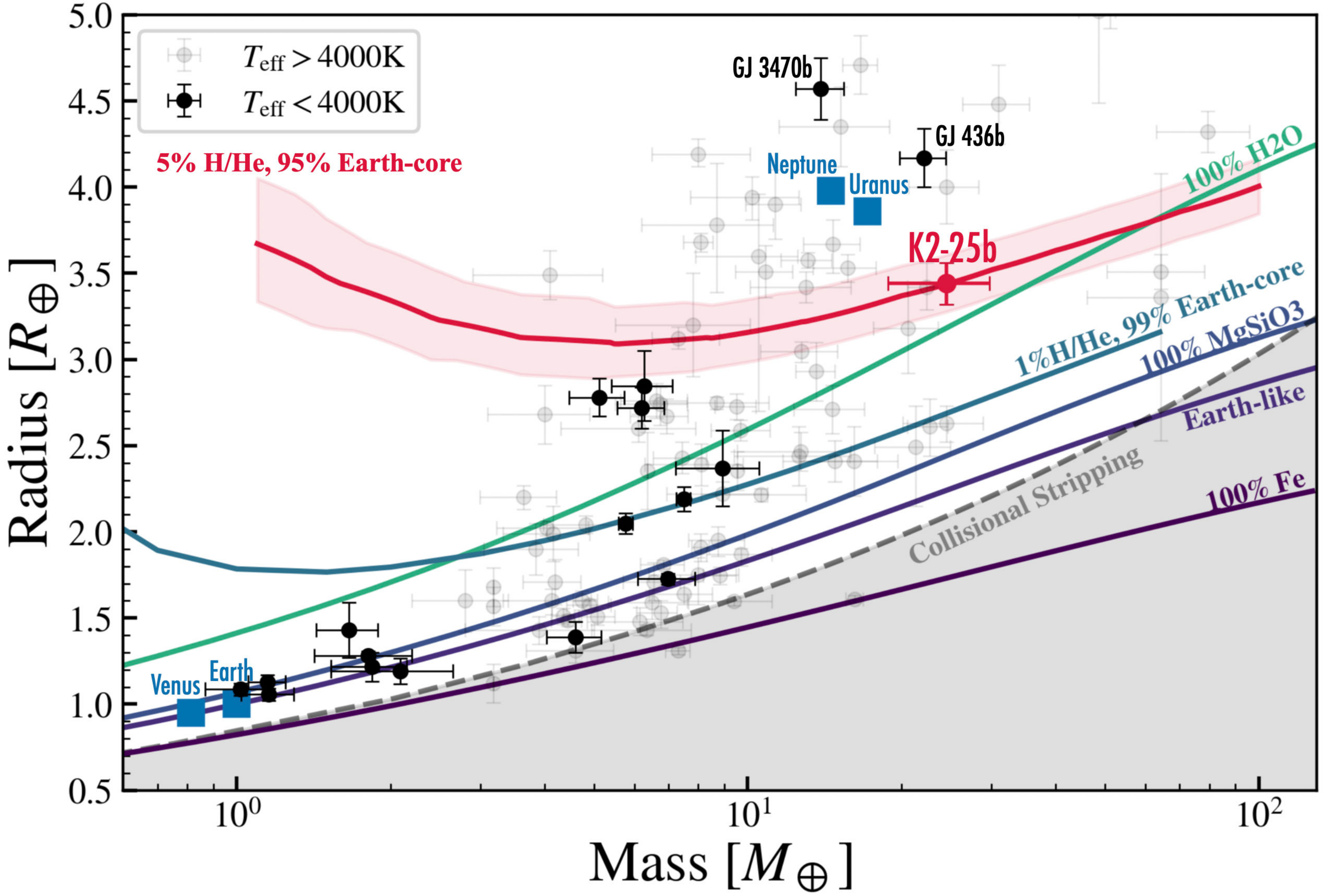}
\caption{K2-25b compared to other similar-sized planets in the exoplanet mass radius plane for M-dwarf planets. M-dwarf planets ($T_{\mathrm{eff}}<4000 \unit{K}$) are denoted with the filled black circles, and planets orbiting hotter stars are shown with the faint grey circles. Blue squares shows solar system planets. The red curve shows our best-fit model from interpolating the two-layer composition model of \cite{lopez2014} assuming a rocky core capped by a H/He envelope, resulting in a H/He mass fraction of $5\%$. The red shaded region shows the associated 68\% credible interval. The other solid lines show the composition models of \cite{zeng2019}. The shaded grey region indicates planets with iron content exceeding the maximum value predicted from models of collisional stripping \citep{marcus2010}. K2-25b is similar in size to the well-studied M-dwarf planets GJ 436b and GJ 3470b.}
\label{fig:mr}
\end{center}
\end{figure*}

\subsection{Improved Ephemeris}
\label{sec:ephemeris}
Figure \ref{fig:ephemeris} shows the improved ephemeris derived by jointly modeling the \textit{K2} and ground-based transits (Model RV2 in Table \ref{tab:planetparams}). In blue we show our ephemeris derived from the K2-data only ($T_{0,K2} = 2457062.5790 \pm 0.0005$ and $P_{K2} = 3.484547 \pm 0.000040$), which agrees well with the ephemeris derived in \cite{mann2016} also from the \textit{K2} data. Our K2-only ephemeris results in a transit timing uncertainty of $\sim$35-40minutes at the start of the JWST era nominally in 2021. Further, we see that our joint-fit ephemeris (shown in black in Figure \ref{fig:ephemeris}) is fully consistent within the 1$\sigma$ errorbars of our K2-only ephemeris. Our joint-fit ephemeris results in a factor of $\sim$150 improvement in the transit timing precision from the \textit{K2} data only, yielding a timing uncertainty of $\sim$$20 \unit{sec}$ at the start of the JWST era nominally in 2021, which will be important for scheduling follow-up observations in the future. Additionally, Figure \ref{fig:ephemeris} shows that our joint-fit is in excellent agreement with the transit ephemeris presented in \cite{thao2020} derived from photometry from \textit{K2}, \textit{Spitzer} and ground-based observations.

\subsection{Composition}
\label{sec:composition}
To compare the possible composition of K2-25b to other planets, in Figure \ref{fig:mr} we plot K2-25b along with other planets in the exoplanet mass-radius plane\footnote{Data retrieved from the NASA Exoplanet Archive \cite{akeson2013} in November 2019}. In Figure \ref{fig:mr}, we only show planets with fractional errors on mass and radius that are better than 25\%, as otherwise, their mass and radius values are consistent with a wide array of planet composition models. The shaded grey region indicates planets with iron content exceeding the maximum value predicted from models of collisional stripping \citep{marcus2010}. The solid lines are theoretical mass-radius curves assuming a constant density from \cite{zeng2019}. From Figure \ref{fig:mr}, we see that at a radius of $R=3.44 R_\oplus$, K2-25b is similar in size to two other well-studied M-dwarf planets: GJ 436b \citep[$4.2 R_\oplus$;][]{maciejewski2014}, and GJ 3470b \citep[$4.57 R_\oplus$;][]{awiphan2016}, which we highlight in orange in Figure \ref{fig:mr}. With a mass of $M=24_{-5.2}^{+5.7} \unit{M_\oplus}$, K2-25b is similar in mass to GJ 436b \citep[$M = 22.1 \unit{M_\oplus}$;][]{maciejewski2014}, but more massive than GJ 3470b \citep[$M = 13.9 \unit{M_\oplus}$;][]{awiphan2016}.  Interestingly, both GJ 436b and GJ 3470b are known to be experiencing substantial atmospheric outflows, resulting in significant atmospheric mass-loss throughout their lifetime (see e.g., \citealt{ehrenreich2015}, and \citealt{bourrier2018b,ninan2019}, respectively). The possibility that K2-25b is experiencing outflows is further discussed in Subsection \ref{sec:transmission}.

There are degeneracies in the composition of planets with radii between $2-4 R_\oplus$: these planets could either have rocky cores with H/He envelopes, or they could be water-rich worlds that contain a significant amount of multi-component water dominated ices/fluids in addition to rock and gas \citep{zeng2019}. As such, from knowing the masses and radii of a planet alone, we can not discern between the different solutions \citep{adams2008}.

Given this known degeneracy, to explore the range of possible compositions for K2-25b, we overplot a number of different compositional growth models in the exoplanet mass-radius plane from \cite{zeng2019}. To place a quantitative estimate on the possible H/He fraction of K2-25b, we modeled the composition of K2-25b assuming a two-layer thermal model consisting of a rocky core and a H/He atmosphere using the models presented in \cite{lopez2014}. In this model, the H/He envelope is the dominant driver of the size of the planet. Assuming this two component model, we linearly interpolated the tables presented in \cite{lopez2014} and together with the posteriors in the observed mass and radius of K2-25b, we estimate that K2-25b has an envelope mass fraction of $5.3_{-0.9}^{+1.2}\%$. The 50th percentile model that best fits the observed mass and radius constraints of K2-25b is shown in the red curve in Figure \ref{fig:mr}, with the $1\sigma$ error intervals denoted by the red shaded region.

Despite these degeneracies, with more mass measurements of planets in young clusters \citep[see e.g.,][]{barragan2019}, we can start to gain further insights onto the exoplanet Mass-Radius distribution as a function of age, which can help place further constraints on planet formation mechanisms and how time-dependent processes such as photo-evaporation sculpt exoplanet Mass-Radius plane.

\begin{figure*}[t!]
\begin{center}
\includegraphics[width=0.95\textwidth]{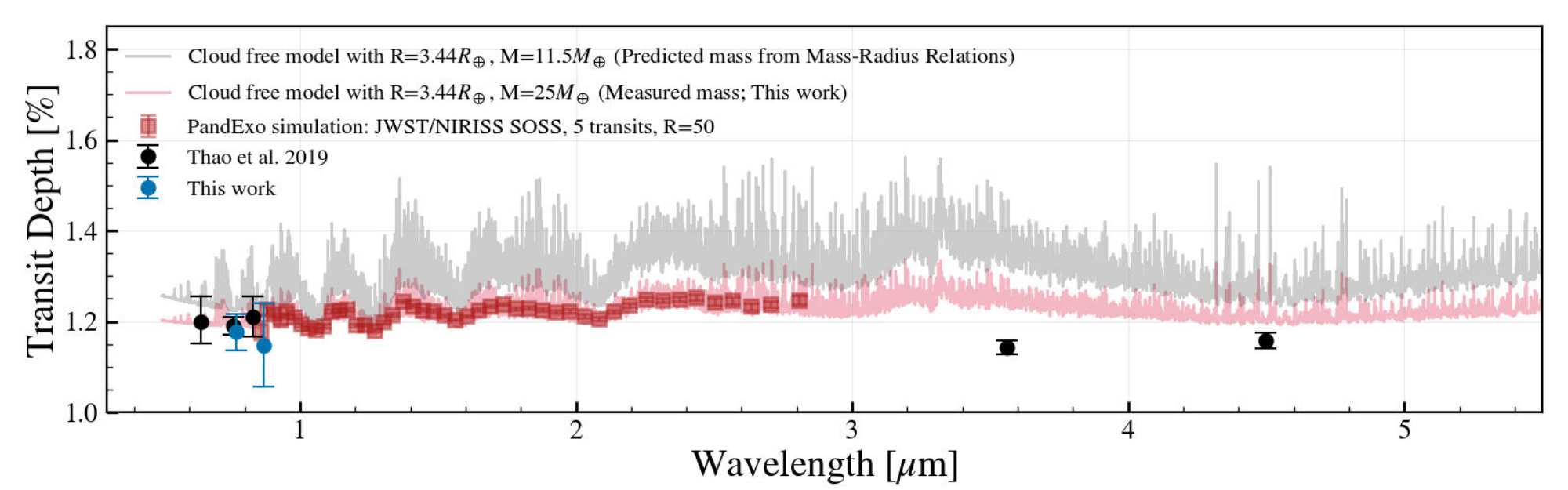}
\caption{Comparison of transit depth as a function of wavelength for K2-25b, from \cite{thao2020} (black points) and from this work (blue points) in the SDSS $i^\prime$ and $z^{\prime}$ bands. We further overlay two expected cloud-free transmission spectra as calculated with the \texttt{pandexo} tool for our measured mass of $25 M_\oplus$ and compare that with the case using the predicted mass of $11.5 M_\oplus$ estimated using the mass-radius relation in the \texttt{Forecaster} package. As expected, our higher mass increases the surface gravity of the planet, which further mutes the expected transmission spectroscopic features.}
\label{fig:transmission}
\end{center}
\end{figure*}

\subsection{Prospects for Transmission Spectroscopy}
\label{sec:transmission}
With its large transit depth, and brightness at NIR wavelengths ($J=11.3$), K2-25b has been mentioned as a prime candidate for transmission spectroscopy \citep[e.g.,][]{mann2016,thao2020}. \cite{thao2020} measured and studied the transit depths of K2-25b in different broad-band filters via precision ground-based photometry, and space-based photometry from \textit{Kepler}, and \textit{Spitzer}. In Figure \ref{fig:transmission}, we plot the transit depth measurements from \cite{thao2020} along with our transit depth measurements in the SDSS $i^\prime$ and SDSS $z^\prime$ bands. As our transit depth measurement presented in Table \ref{tab:planetparams} assumed a single transit depth for all bands, to measure these transit depths, we conducted two separate transit fits using the \texttt{Juliet} program: first, we jointly modeled our five SDSS $z^\prime$ band transits, and second, we jointly modeled our four SDSS $i^\prime$ band transits. From Figure \ref{fig:transmission}, we see that our transit depths in these bands agree well with the optical transit depth measurements in \cite{thao2020}. However, as is seen in Figure \ref{fig:transmission}, the infrared \textit{Spitzer} transits from \cite{thao2020} are statistically smaller than the optical transits. As is detailed in \cite{thao2020}, this could potentially be due to starspots causing the transit depths to be less deep. However, like they argue, the impact of starspots on the transit depth in the NIR is likely lower than in the optical. Further, no clear starspot crossing events are observed in their data, the data studied in \cite{kain2019} (although \cite{kain2019} mention a few candidate starspot crossing events) or in the transits presented here. As concluded by \cite{thao2020}, while spots can have an impact on the overall transmission spectrum, it is unlikely that starspots alone could cause the difference in the NIR transit depths. One possible explanation, as mentioned in \cite{thao2020} is that K2-25b has a predominantly cloudy or hazy atmosphere, causing a flat transmission curve as a function of wavelength.

\cite{thao2020} use the parametric mass-radius relation from \cite{wolfgang2016} to estimate the most likely mass for K2-25b of $M=13M_\oplus$, and consider different transmission models assuming planet surface gravities of $6 \unit{m\:s^{-2}}$, $9 \unit{m\:s^{-2}}$, and $12 \unit{m\:s^{-2}}$, corresponding to planet masses of $8 M_\oplus$, $11.5 M_\oplus$, and $15 M_\oplus$, respectively. All of these mass estimates are lower than our measured mass of $M=24.5_{-5.2}^{+5.7} M_\oplus$, which corresponds to a surface gravity of $g = 20.3 \pm 4.7 \unit{m\:s^{-2}}$. As an independent test, we also predicted a mass of $M=11.5_{-4.8}^{+9.1}M_\oplus$ using the mass-radius relation in the \texttt{Forecaster} package from \cite{chen2017}. In Figure \ref{fig:transmission}, we compare the expected transmission spectrum as calculated with \texttt{pandexo} package \citep{batalha2017pandexo} for both the predicted mass of $11.5 M_\oplus$, and our measured mass of $25 M_\oplus$. As expected, we see that the transmission features of K2-25b are muted in the more massive case, as the increased mass increases the surface gravity of the planet. This can also be seen if we calculate the Transmission Spectroscopy Metric (TSM) defined in \cite{kempton2018}: using the predicted mass of $11.5 \unit{M_\oplus}$ we obtain a TSM=143, while if we use our mass estimate, we estimate a lower of $\mathrm{TSM}=66_{-14}^{+21}$. Our larger mass measurement which mutes the expected transmission features of K2-25b could further help---at least partially---explain the flat transmission features observed by \cite{thao2020}. 

Despite our larger mass estimate causing the expected transmission features to be muted, the cloud-rich or hazy atmosphere scenario suggested in \cite{thao2020} is still a likely possibility. In fact, this would conform with the rising statistical trend that colder planets preferentially show muted and/or flat features in their transmission spectra \cite{crossfield2017}. However, to confidently quantify the exact amplitude of the transmission features---which strongly depend on the mass of the planet, which is now known---we argue that additional observations with JWST will be valuable to further confidently rule out or confirm a flat transmission spectrum (particularly the information-rich $1-2\unit{\mu m}$ region). As an example, in Figure \ref{fig:transmission}, we overplot the expected S/N of JWST/NIRISS in Single Object Slitless Spectroscopy (SOSS) mode using the \texttt{gr700xd} grism after 5 transit observations binned to a resolving power of $R=50$, showing that even for our large mass of $24.5M_\oplus$, JWST/NIRISS should have the sensitivity to confidently discern between a clear or cloudy atmosphere. Gaining further insights into K2-25b's atmosphere will be particularly valuable, as this will allow us to put constraints on the atmospheric constituents of this relatively young planet, giving key insights into the atmospheric composition of adolescent planets and planetary atmospheres as a function of time. We note that \cite{wang2019} suggest that young low-density 'super-puff' planets---planets with mean density $\rho < 10^{-1} \unit{g/cm^3}$---could be susceptible to extreme hydrodynamic mass loss which can carry large numbers of small dust particles to high altitudes, which in turn can create featureless transmission spectra. However, with K2-25b having a bulk density of $\rho = 3.3 \pm 0.8 \unit{g/cm^3}$, which is substantially larger than the cutoff for 'super-puff' planets, this scenario is unlikely to be the case.

\begin{figure*}[t!]
\begin{center}
\includegraphics[width=0.9\textwidth]{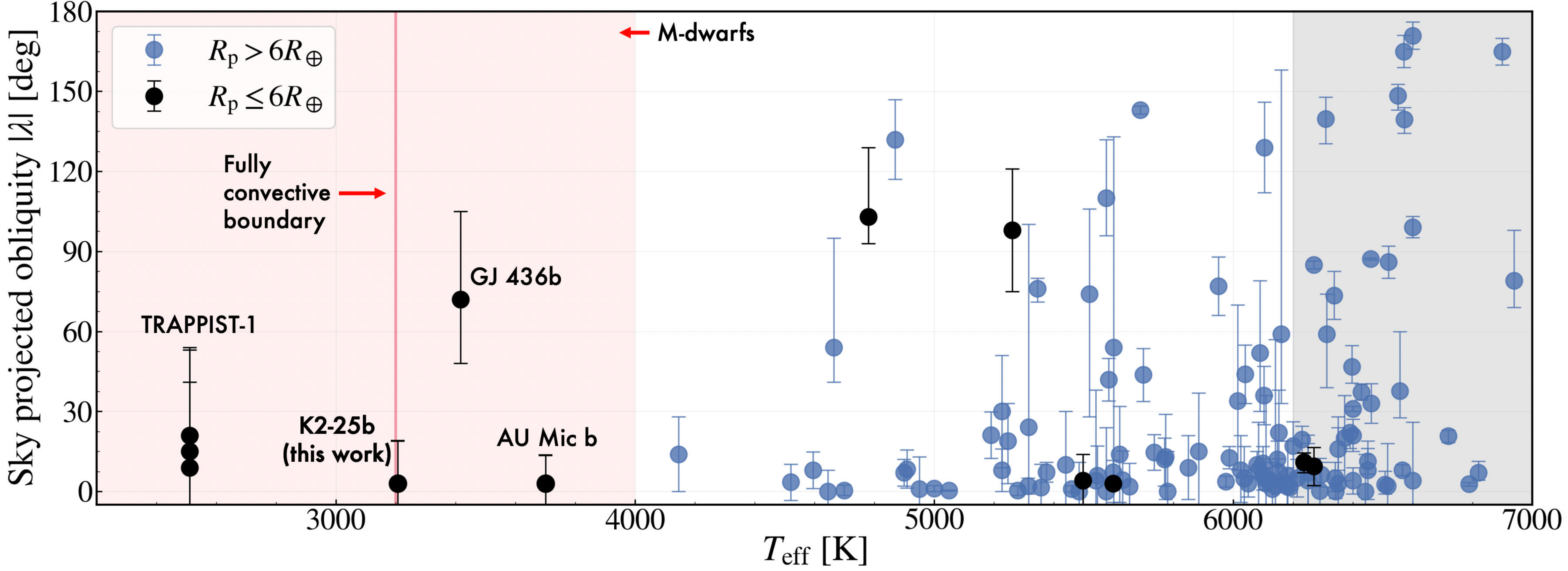}
\caption{Currently available sky-projected obliquity constraints $\lambda$ obtained from the TEPCAT database \citep{southworth2011} as a function of stellar effective temperature. Small planets ($R < 6 R_\oplus$) are highlighted in black, and larger planets in blue. The grey area shows stellar hosts past the Kraft-break \citep{kraft1967}, where stars loose their outer convective layers and become fully-radiative. M-dwarf systems are shaded in red, with the onset of fully-convective stars shown with the red solid line. Currently only four M-dwarf planetary systems have their obliquity measured via the RM effect: GJ 436b \citep{bourrier2018}, TRAPPIST-1 \citep{hirano2018}, AU Mic b \citep{addison2020,hirano2020mic,palle2020}, and K2-25b (this work).}
\label{fig:psis}
\end{center}
\end{figure*}

\subsection{Obliquity and Orbital Alignment}
\label{sec:alignment}
Figure \ref{fig:psis} compares the sky-projected obliquity of K2-25b to currently available sky-projected obliquity measurements of other planetary systems from the TEPCAT\footnote{TEPCAT database: \url{https://www.astro.keele.ac.uk/jkt/tepcat/obliquity.html}} database \citep{southworth2011}, where M-dwarfs are shaded in red and the onset of fully-convective stars is highlighted with the red solid line. Our obliquity constraint of K2-25b marks the fourth obliquity constraint of an M-dwarf planet system via the RM effect, the other being GJ 436b as measured by \cite{bourrier2018}, AU Mic b as measured by \cite{addison2020}, \cite{hirano2020}, and \cite{palle2020}, and TRAPPIST-1 as measured by \cite{hirano2020}. At an age of 600-800 MYr, K2-25b has an intermediate age among these systems: TRAPPIST-1 has an estimated age of $7.6\pm 2.2$ GYr \citep{burgasser2017}, GJ 436 an age between 4-8 GYr \citep{bourrier2018}, and AU Mic is the youngest with an age of 22 MYr. Interestingly, both K2-25b and GJ 436b have eccentric orbits ($e=0.43\pm0.05$ and $e=0.1616 \pm 0.004$, respectively) and are similar in size, but they show different orbital architectures: GJ 436b observed to have a misaligned orbit while K2-25b is observed to have an aligned orbit. This could be suggestive of a different formation and subsequent dynamical history. \cite{bourrier2018} suggest that von Zeipel-Lidov-Kozai migration induced by a candidate perturber could explain both GJ 436b's eccentricity and misaligned orbit. However, as we do not see that K2-25b is heavily misaligned this disfavors von Zeipel-Lidov-Kozai cycles acting on the planet. We further discuss potential formation pathways for K2-25b in Subsection \ref{sec:pathways}.

With only a few obliquity measurements of M-dwarf planetary systems published via the RM effect, the orbital architectures of individual M-dwarf planetary systems remain relatively unexplored. However, we note that statistical studies suggest that planets orbiting cooler planet hosts have orbits that are on average better aligned to their stellar equators than planets orbiting hotter stars. In comparing the rotation distribution of Kepler Objects of Interests (KOIs) hosting transiting planet candidates to a control sample of single stars without transiting planet candidates, \cite{mazeh2015} showed that hotter stars ($T_{\mathrm{eff}} > 6000 \unit{K}$) show on average lower amplitude photometric modulations suggesting a broad distribution of obliquities, while cooler KOIs ($3500 \unit{K} < T_{\mathrm{eff}} < 6000 \unit{K}$) showed on average higher amplitude modulations, suggesting well-aligned systems. This result is in broad alignment with the findings of \cite{winn2010b} \cite{schlaufman2010}, and \cite{albrecht2012}, that hotter stars ($T_{\mathrm{eff}} > 6250 \unit{K}$) hosting close-in gas giants show a broad distribution of obliquities and misalignments, while cooler stars ($T_{\mathrm{eff}} < 6250 \unit{K}$) tend to host well-aligned systems. We note that neither of these studies encompassed mid-to-late M-dwarfs. In general, across the M-dwarf spectral type---and especially for mid-to-late M-dwarfs---the occurrence of gas giants is lower than seen around FGK stars \citep{johnson2010,dressing2015,hardegree2019}. By further studying the orbital architectures of M-dwarf systems, which have a lower occurrence of massive planets, we can gain further insights into the role massive planets play in sculpting the orbital architectures across different exoplanet host stars. The advent of precise NIR spectrographs, such as HPF \citep{mahadevan2012,mahadevan2014}, the Infrared Doppler Instrument \citep[IRD; ][]{kotani2018}, CARMENES \citep{quirrenbach2018}, SPIROU \citep{artigau2014}, NIRPS \citep{wildi2017}, GIANO-B \citep{claudi2018}, and red-optical spectrographs on large telescopes such as MAROON-X \citep{seifahrt2016}, ESPRESSO \citep{pepe2018}, and KPF \citep{gibson2016}, is opening the doors to the ensemble study of obliquities of M-dwarfs.

\subsection{Search for Additional non-transiting Planets in the HPF RVs}
\label{sec:planetc}
To place an upper limit on a potentially non-transiting second planet in the system, we performed an additional two-planet RV fit of the HPF RVs using \texttt{juliet}. As there is no obvious signs of another transiting planet in the system in the \textit{K2} data, we only considered the HPF RVs for this fit. We assumed a two planet model along with a quasi-periodic Gaussian Process to account for correlated noise due to stellar activity at the known rotation period of the star. We placed Gaussian priors on the orbital parameters of K2-25b derived from the joint fit of the RVs and the photometry, and used the same priors we placed on the Gaussian Process hyperparameters as listed in Table \ref{tab:priors1} for fits RV1 and RV2. For the hypothetical planet c, we placed broad priors on the period (Jeffreys prior from 0.5days to 50days), time of conjunction (modeled as a transit midpoint uniform from 50days before the first RV point to 50days after the last RV point), eccentricity (uniform from 0 to 0.95), argument of periastron (uniform prior from $0^\circ$ to $360^\circ$), and RV semi-amplitude (uniform from 0 to 500m/s). To check for evidence of a long-term slope, we also added a radial velocity slope with broad uniform priors.

\begin{figure}[t!]
\begin{center}
\includegraphics[width=1\columnwidth]{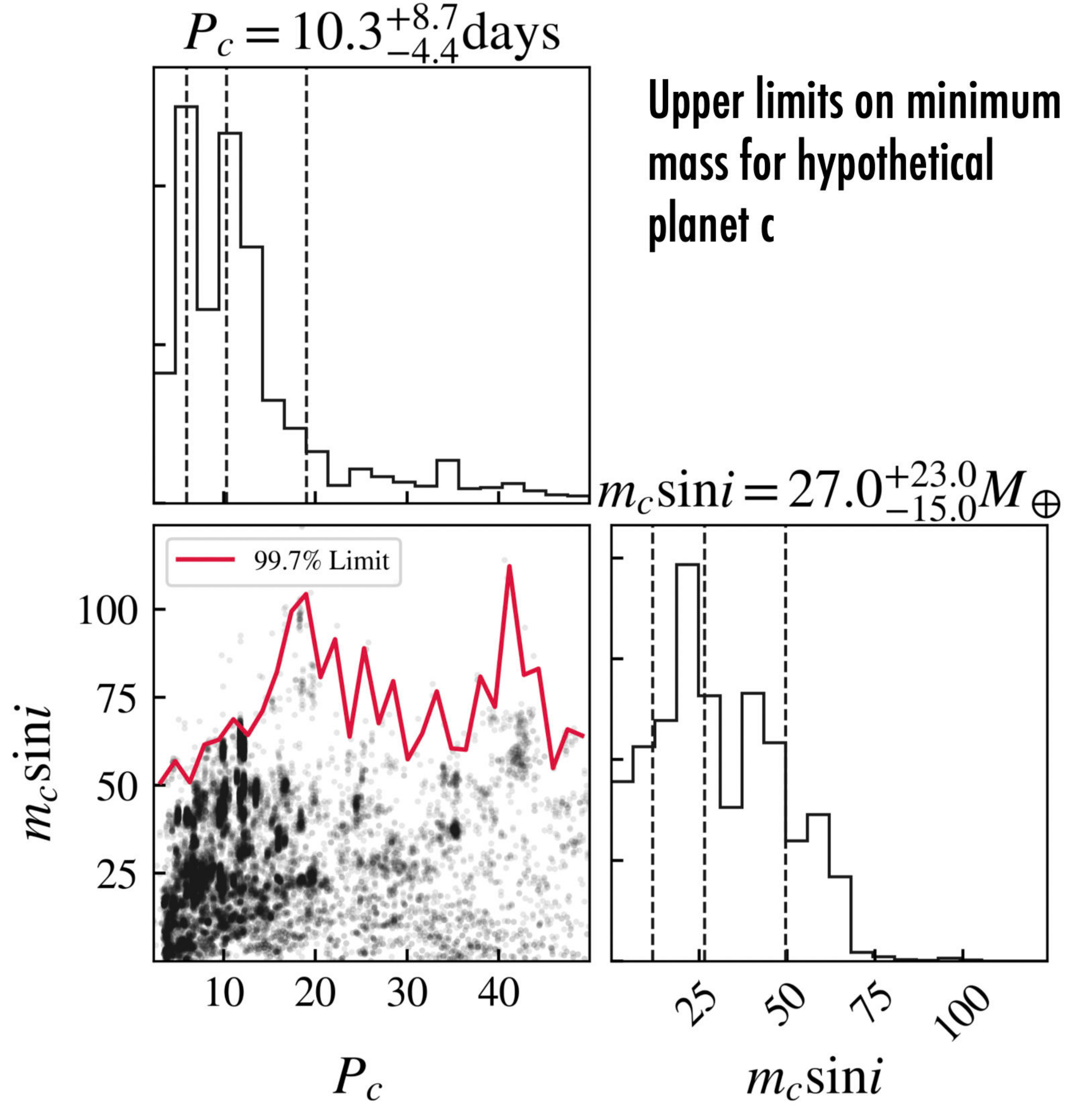}
\caption{Posterior constraints on a hypothetical non-transiting planet c in the HPF RVs in mass ($m_c \sin i$) and orbital period ($P_c$) space. We see no obvious evidence for another massive planet in the HPF RVs given the broad posterior constraints. The red curve shows a running 99.7\% percentile upper limit ($3\sigma$) on the mass as a function of period. Across the full period range considered, we use the posteriors to place an upper mass limit of $82M_\oplus$ at 99.7\% confidence for periods between 0.5 and 50days.}
\label{fig:planetc}
\end{center}
\end{figure}

Figure \ref{fig:planetc} shows the constraint on the orbital period and mass ($m\sin i$) of the hypothetical planet c. From this, we see that no obvious preferred solution is found. From the posteriors, we place an upper limit on the mass of a possible secondary planet of $m \sin i < 82 M_\oplus$ at 99.7\% confidence ($3\sigma$) for periods between 0.5days to 50days. The red curve in Figure \ref{fig:planetc} shows a running 99.7\% upper limit on the mass as a function of period for smaller period bins. We further note that our constraint on an additional radial velocity slope is consistent with zero slope within the $1\sigma$ uncertainties. We additionally compared the log-evidence values we obtained from our \texttt{juliet} fit for the 2-planet model (2 planets, a GP to account for stellar activity, and an RV slope), to a null model assuming only K2-25b in the system (1 planet and GP to account for stellar activity). In doing so, the two planet model had a log-evidence value of $\ln(Z) = -183.7 \pm 0.5$, while the null one planet model had a log-evidence value of $\ln(Z) = -170.6 \pm 0.1$, where we have reported the mean and standard deviation of 6 independent runs of each model to get an accurate estimate of the spread in log-evidence values. We see that the one planet model is statistically favored with a higher evidence of $\Delta \ln(Z) = 13.1$. With the current RV data in hand, we rule out massive companions with masses $m \sin i > 82 M_\oplus$ at 99.7\% confidence with periods between 0.5 and 50 days, and conclude that we do not have sufficient evidence to claim another small short period planet in the system.

\subsection{Formation and Subsequent Evolution}
\label{sec:pathways}
From our detailed constraints of K2-25b's planetary properties, including constraints on its mass, eccentricity, volatile content, and obliquity of the host star, allows us to put informative constraints on potential formation scenarios, which we discuss below.

\subsubsection{Potential Formation Scenarios}
\label{sec:formation}
In the core accretion model of planet formation, planetesimals collide to form protoplanetary cores, which then attract a gaseous envelope \citep[e.g.,][]{bodenheimer1986,pollack1996}. If the planet core becomes sufficiently massive---after reaching a critical mass $M_{\mathrm{crit}}$---before the gaseous protoplanetary disk dissipates, the protoplanetary core can enter a phase of runaway gas accretion where the planet attracts a massive gaseous envelope. Although the critical core mass is typically quoted as 10 Earth masses, it can vary by orders of magnitude depending on the disk conditions and planetesimal accretion rate \citep[see e.g.,][]{rafikov2006}. The gaseous envelope is accreted onto the planetary core from the gaseous component of the protoplanetary disk, which only remains present for a few million years around young stars \citep{williams2011,ribas2015}, suggesting that core formation must happen within this timeframe. However, traditional core accretion models suffer from predicting formation timescales for gaseous planets that are much larger than the disk dissipation timescale \citep{dodson2009,rafikov2011}.

In the pebble accretion model of planet formation, small millimeter to centimeter sized pebbles accrete onto a planetary core \citep{lambrechts2012,lambrechts2014,lambrechts2017}. These pebbles are marginally coupled to the nebular gas on orbital timescales, creating sufficient gas drag to enable efficient core formation possible within the disk-dissipation timescale even at large orbital separations \citep{lambrechts2012}. The pebble accretion model predicts that pebble accretion terminates when the planet reaches the 'pebble isolation mass', $M_{\mathrm{iso}}$, the mass when the gravity of the core is strong enough to open a gap in the disk that hinders further accretion of pebbles \citep{lambrechts2014}. If the pebble isolation mass is greater than the critical core mass and the critical core mass is reached before the disk dissipates of gas, the massive core can rapidly accrete gas to form a gas giant. However, if the pebble isolation mass is not reached and/or if it is reached after most of the disk dissipated and/or it is below the critical core mass, the planet core consists primarily of accreted icy pebbles with a minimal H/He atmosphere. Although the exact value of the pebble isolation mass depends strongly on a number of uncertain disk properties including the aspect ratio, viscosity, local disk structure, and the pebble size, the pebble isolation mass for solar-type stars at 5 AU has been estimated to be $10-20 \unit{M_\oplus}$ \citep{lambrechts2014}, although larger values are also possible depending on the assumed disk properties \citep[see e.g.,][]{bitsch2018}. The pebble isolation mass increases with the orbital semi-major axis proportional to $a^{3/4}$ \citep{lambrechts2014}, where $a$ is the semi-major axis of the planetary core. This suggests that planets with massive dense cores can not form too close to the host star where the pebble isolation mass is smaller. Further, the pebble isolation mass is expected to decrease for less massive later-type stars according to the following formula from \cite{liu2019},
\begin{equation}
M_{\mathrm{iso}} = 25 \left( \frac{M_*}{\mathrm{1M_\odot}}\right)^{4/3} M_\oplus,
\label{eq:iso}
\end{equation}
where $M_*$ is the mass of the host star, and assuming that the planet reaches the isolation mass at the ice-line distance. Assuming K2-25b formed at the ice-line and subsequently migrated further in, we calculate an expected isolation mass of $\sim$$5M_\oplus$.

At its currently observed mass of $24.5 \unit{M_\oplus}$ with a thin 5\% H/He envelope, K2-25b is nominally at odds with the predictions of the core-accretion and pebble-accretion models if it reached its final mass during the full gas disk stage beyond the ice line under certain disk conditions and assumed planetesimal accretion rates. From the core-accretion model, at its currently observed mass we would have predicted K2-25b to have experienced runaway gas accretion and to have formed a gas giant, while we instead infer today only a thin H/He envelope of $\sim$5\%. From the viewpoint of the pebble-accretion model, we would expect that K2-25b would have reached a nominal maximum core mass mass that is closer to the isolation mass of $\sim5M_\oplus$ that we estimated from Equation \ref{eq:iso} if it formed beyond the ice line. We again acknowledge that that the exact value of the pebble isolation mass is uncertain as it depends strongly on a number of gas disk properties for K2-25b which are not well known.

To arrive at the presently observed mass, and to explain K2-25b's low inferred H/He content and moderate eccentricity, a possible scenario is that K2-25b grew its mass through the merging of planetary cores. The merging of icy cores has been postulated to explain the observed bimodality in the mass distribution of Neptunes between 2-4$R_\oplus$ \citep{zeng2019}. Although collisional growth could happen at K2-25b's current orbital location after e.g., in-situ formation or disk-driven migration, exciting the planet's current high eccentricity could be more easily excited at larger orbital distances (we discuss its high eccentricity and migration pathways further below). If the seed cores were composed primarily of water ices, \cite{marcus2010} show that the collisions of two icy cores tend to stick yielding a final core of doubled mass. During collisions, it is possible that substantial fractions of H/He envelopes get stripped away \citep[see e.g.,][applicable for masses lower than Neptune]{inamdar2015}, which could mean its atmospheric fraction was somewhat higher in the past (though still not a gas giant). 

Lastly, a possible scenario is that K2-25b formed in-situ close to its current observed orbit close to the host-star \citep[e.g.,][]{batygin2016}, followed by dynamical interactions to explain its current eccentricity. In-situ formation has been shown to be able to describe the observed compositional and orbital diversity of super-Earths and mini-Neptunes through inherent variations in the initial formation conditions in the disk \citep{macdonald2020}. To avoid runaway accretion in an in-situ formation scenario, as \cite{lee2014,macdonald2020} and others have argued, the accretion could happen at a later stage when the disk is partially depleted, then grow through giant impacts, and accrete a gaseous envelope in a depleted nebula. However, a challenge for this formation scenario is that in-situ formation generally predicts eccentricities that are lower than K2-25b's currently observed eccentricity \citep[see e.g.,][]{macdonald2020}.

\subsubsection{Potential migration histories and eccentricity}
From the viewpoint of core and/or pebble-accretion, it is possible that K2-25b formed at large orbital distances where these models predict the creation of more massive planetary cores which later collided together to assemble a more massive planet. If K2-25b formed further out, then K2-25b must have needed to migrate closer to the star to arrive at its current short $P=3.48 \unit{day}$ orbit with its moderate eccentricity of $e=0.428$. As such, K2-25b could still be in the process of migrating to an even shorter orbit through tidal migration and could represent a precursor of hot Neptunes seen around older stars. There are two major possibilities for migration of K2-25b: disk-driven migration, and high-eccentricity migration.

Disk-driven migration \citep[see e.g.,][and references therein]{kley2012} relies on the exchange of angular momentum between the disk and the planet through mutual gravitational interactions within the plane of the disk. These interactions tend to migrate planets from long period orbits to shorter period orbits while damping eccentricities and thus resulting in circular orbits within the timescale of disk-dissipation \citep{kley2012}. Therefore, to explain K2-25b's current moderate eccentricity, a round of dynamical interactions would be needed. However, K2-25b's eccentricity is much larger than we expect from eccentricity excitation following migration \citep{petrovich2014}.

\begin{figure*}[t!]
\begin{center}
\includegraphics[width=0.9\textwidth]{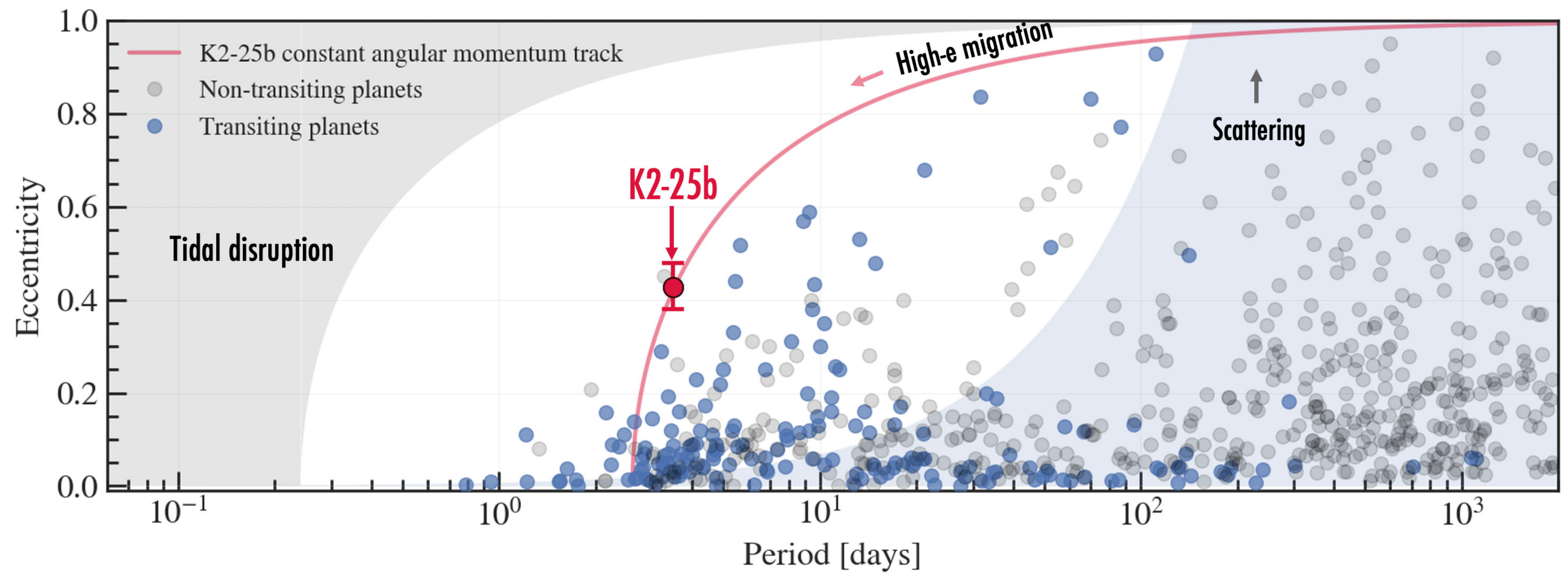}
\caption{Eccentricity as a function of orbital period for known exoplanet systems: grey points show known non-transiting systems, and the blue points show known transiting planets. The blue region shows the region where scattering tends to leads to eccentricity excitation rather than collisions (calculated assuming K2-25b's current mass of $25 M_\oplus$). Seed planetary cores could have formed at these orbital distances, and then scattered through gravitational interactions to a high eccentricity orbit. Subsequently, K2-25b could have migrated closer to its host star via tidal interactions, arriving at its current orbit. The red curve shows the track of constant angular momentum for K2-25b, showing a possible migration pathway. Only systems with eccentricity errors less than $<0.1$ are shown.}
\label{fig:eccentricity}
\end{center}
\end{figure*}

Another possible scenario is that K2-25b arrived at its present orbit via high-eccentricity migration, a process which appears to be the dominant process for generating short-period giant planets \citep{dawson2018}, and has been used to explain the migration process for most observed warm giants with $e>0.4$ \citep{petrovich2016}. High eccentricity migration can often be approximated as a two step process: reducing the planet's orbital angular momentum, and then reducing its energy \citep{dawson2018}. During the first step, a perturber extract orbital angular momentum from the planet by perturbing it into a highly elliptical orbit. In the second step, the planet tidally dissipates its orbital energy through interactions with the central star. If so, what originally excited K2-25b's eccentricity? Several theories have been proposed to explain the original excitation of eccentricities, including planet-planet scattering and/or merging events \citep{rasio1996,chatterjee2008,juric2008}, secular chaos \citep{wu2011}, and stellar flybys \citep[e.g.,][]{kaib2013}. Additionally, secular interactions---e.g., via the von Zeipel-Lidov-Kozai mechanism from a widely separated perturber \citep{naoz2016,ito2019}, or coplanar secular interactions \citep{petrovich2015}---can trigger high eccentricity migration. However, significant spin-orbit misalignment and retrograde motion can result from the von Zeipel-Lidov-Kozai mechanism \citep[e.g.,][]{storch2017,bourrier2018}. Therefore, to explain the observed low obliquity of K2-25b, we argue that this mechanism is less likely the cause for K2-25b's moderate eccentricity. Instead, to explain K2-25's low inferred obliquity, planet-planet scattering events and/or coplanar secular interactions are more likely mechanisms as they can produce systems with high eccentricities but low inclinations \citep[see e.g.,][]{chatterjee2008,petrovich2014,petrovich2015}.

Figure \ref{fig:eccentricity} compares the orbital eccentricities of exoplanets as a function of orbital period\footnote{Data obtained from the NASA Exoplanet Archive \citep{akeson2013}.}, showing that K2-25b is among select few planets with moderate eccentricities ($e>0.4$) at short orbital periods ($<10 \unit{days}$). The region to the right of the grey curve in Figure \ref{fig:eccentricity} is the region where planet-planet scattering can take place to excite eccentricities, and not preferentially cause collisions\footnote{Calculated assuming K2-25b's current mass of $\sim25M_\oplus$ using Equation 10 in \citep{dawson2018} which compares the escape velocity of the planet's surface to the escape velocity from the star at the planet's semi-major axis.}. As discussed above, a possible formation scenario for K2-25b is that initial planet seed cores at the isolation mass formed via pebble accretion at long orbital periods, which then scattered via dynamical interactions to high eccentricities. These highly eccentric orbits then potentially led to orbit crossings and subsequent collisions resulting in the dense planet we see today, which could then be currently migrating towards a shorter circular orbit through tidal interactions with the star. Figure \ref{fig:eccentricity} highlights a nominal migration track assuming a track of constant angular momentum. Extrapolating this track suggests a fully circularized orbit with a period of $\sim$2.5 days (see Figure \ref{fig:eccentricity}). 

Using tidal dissipation theory, we can gain further insight into the plausibility of this formation scenario. Tidal dissipation mechanisms vary strongly with the internal structure of the planet \citep{goldreich1966,guenel2014}, with higher tidal quality factors $Q$ seen for the gas giant planets than the denser rocky planets in the solar system \citep{goldreich1966}\footnote{The tidal quality factor $Q$ is a factor quantifying the degree of tidal dissipation in gravitational systems and is inversely proportional to the degree of dissipation.}. Using the equations in \cite{jackson2009}, \cite{kain2019} estimate a circularization timescale of 410 MYr for K2-25b, assuming a modified tidal quality factor\footnote{The modified tidal quality factor $Q^\prime$ is defined as $3Q/2k_2$, where $k_2$ is the Love number of degree 2 \citep{goldreich1966} for the system.} of $Q_*^\prime = 10^5$ for the host star, and $Q_p^\prime = 5 \times 10^4$ for K2-25b which they selected based on Neptune's most likely $Q_p^\prime$ value from \cite{zhang2008}. As further noted by \cite{kain2019}, the exact value of the tidal circularization timescale for K2-25b scales directly with $Q_p^\prime$. Assuming the same tidal quality factors for the planet and the star as \cite{kain2019}, but using the planetary mass, radius, and $a/R_*$ values derived in this work, we estimate a circularization timescale of 306 MYr, which is slightly lower but broadly consistent with their value. We formally require a tidal quality factor of $Q_p^\prime > 1\times10^{5}$ to achieve a tidal circularization timescale consistent with or longer than the age of the K2-25 system (650-800MYr). If the tidal quality factor of K2-25b is indeed $Q_p^\prime = 1 \times 10^5$ or larger, this could explain the moderate eccentricity of K2-25b we observe today, suggesting that at its current age of 650-800 MYr, K2-25b has not had sufficient time to circularize its orbit. Although larger than Neptune's quality factor, we note that it has been hypothesized that the interior structures of close-in Neptune-sized planets may differ from those of the more distant ice giants in our solar system; in particular, \cite{morley2017} reported a high dissipation factor for GJ 436b of $Q_p^\prime = 10^5-10^6$, which has been theorized to help explain the moderate eccentricity $e=0.16$ observed for GJ 436b. 

If however, we assume a lower tidal quality of $Q_p^\prime = 5\times10^3$---a representative tidal quality factor between the tidal quality factors of the rocky planets and the gas giants in the Solar system \citep{goldreich1966}---we obtain a circularization timescale of 30 MYr, which is substantially shorter than the age of the system (650-800 MYr). If the tidal quality factor is indeed this low, K2-25b would likely require additional ongoing eccentricity excitation to account for its currently observed moderate eccentricity. Although ongoing eccentricity excitation via Kozai-Lidov cycles is less likely given the low measured obliquity, coplanar secular excitation---as has been explored by \cite{batygin2009} to account for the moderate eccentricity of GJ 436 b---is a possibility. However, as is discussed in Section \ref{sec:planetc}, we do not see any clear evidence for a second massive planetary companion in the HPF RVs, although more precision RVs are needed to confidently rule out the presence of smaller planets that could be present. Another more fine-tuned possibility that we can not rule out from the available data, is that K2-25b was only recently (around 30 MYr ago) excited to an eccentric orbit far from the star, and could then still be in the process of circularizing its orbit.

With a single transiting planet seen in the \textit{K2} data, K2-25b is consistent with the trend observed in \cite{vaneylen2019}, that single transiting planets tend to have higher eccentricities than planets in systems with multiple transiting planets, and the trend that \cite{dong2018} observed that hot-Neptunes---planets with $2-6 R_\oplus$ and an orbital period $P < 10 \unit{days}$---are most commonly found in single transiting systems. Further, \cite{petigura2017} note that more massive sub-Saturns tend to have moderately eccentric orbits and orbit stars without other detected planets. By detecting more young Neptune-sized systems, we can compare their observed properties (e.g., eccentricities and obliquities), and compare them to the predictions from different formation and migration mechanisms to start to establish a clearer picture on how short period Neptune-sized planets originate.

\subsection{Independent Analysis of K2-25b by IRD}
During the preparation of this manuscript, we became aware that Gaidos et al. 2020 (submitted) performed a complimentary analysis of the K2-25b system to constrain its obliquity using the Infrared Doppler Instrument (IRD) on the 8.2m Subaru Telescope. Although the submissions of these complementary studies were coordinated between the groups, the data analyses and interpretations were performed independently.

\section{Summary} 
\label{sec:summary}
We present the first mass and obliquity constraint for the young Neptune-sized planet K2-25b orbiting its M4.5-dwarf host star in the Hyades cluster. Given its known age and well characterized orbital parameters, K2-25b is a benchmark system to study M-dwarf planet formation and subsequent dynamics, giving us further insights into the formation and migration mechanisms that produce other hot Neptune exoplanets.

To characterize the planet properties, we jointly fit the available \textit{K2} photometry along with precision diffuser-assisted ground-based photometry obtained with the Engineered Diffuser on the ARCTIC imager on the 3.5m telescope at Apache Point Observatory and the newly installed Engineered Diffuser on the Half-Degree Imager (HDI) on the 0.9m WIYN Telescope at Kitt Peak Observatory, along with precision out-of-transit NIR RVs from the Habitable-zone Planet Finder (HPF) spectrograph at the 10m Hobby-Eberly Telescope (HET) at McDonald Observatory. We see clear evidence for starspot activity in both the \textit{K2} data, and the HPF RVs and associated activity indicators. Jointly fitting the available photometry and RVs suggests a best-fit radius of $R=3.44\pm0.12 R_\oplus$, an eccentric orbit of $e=0.41 \pm 0.05$, and a mass of $M = 24.5_{-5.2}^{+5.7}\unit{M_\oplus}$. We tested the robustness of our HPF mass measurement by conducting injection-and-recovery tests in synthetic RV streams. Using our radius and mass constraints, and assuming a two-component composition model of a rocky core enshrouded by a thin H/He envelope, we obtain a H/He envelope mass fraction of $5\%$. No obvious long-period massive companion is detected in the HPF RV data and continued precise RV monitoring is required to confidently detect or exclude such a companion.

To constrain the obliquity of the system, we present three Rossiter-McLaughlin effect observations of K2-25b obtained with HPF, yielding a sky-projected obliquity constraint of $\lambda = 3 \pm 16^\circ$. Using our constraint for the stellar inclination derived from the stellar radius and our $v \sin i_* = 8.8 \pm 0.6 \unit{km/s}$ constraint from the high resolution HPF spectra, we obtain a true 3D obliquity of $\psi = 17_{-8}^{+11\:\circ}$. Our obliquity and eccentricity constraints paints a picture of a well-aligned, but eccentric system.

With precisely determined age and orbital parameters, we discuss a few possible formation scenarios for K2-25b. If K2-25b reached its current mass during the gas disk phase beyond the ice line, K2-25b would be at odds with the predictions of core and pebble-accretion models---with certain assumptions about the disk properties and planetesimal accretion luminosity---as those models predict that K2-25b should have experienced runaway gas accretion resulting in a gas giant planet. To explain its currently observed mass, we surmise that K2-25b could be the product of planet merging events of smaller planetary cores to produce a more massive planet. Such a dynamical environment could have excited K2-25b into an eccentric orbit, and K2-25b could be in the process of migrating to a shorter period orbit through tidal interactions with the host star. To explain K2-25b's current moderate eccentricity from tidal circularization theory, we place a lower limit on the tidal quality factor of $Q_p^\prime \sim 10^5$, corresponding to a circularization timescale consistent with the age of the system. This tidal quality factor is higher than the tidal quality factor for Neptune, which suggest that K2-25b's internal structure could be different than that of the small gas-giants (Uranus, Neptune) in the Solar System.

\acknowledgments
We thank the anonymous referee for a thoughtful reading of the manuscript, and for useful suggestions and comments which made for a clearer and stronger manuscript. This work was partially supported by funding from the Center for Exoplanets and Habitable Worlds. The Center for Exoplanets and Habitable Worlds is supported by the Pennsylvania State University, the Eberly College of Science, and the Pennsylvania Space Grant Consortium. This work was supported by NASA Headquarters under the NASA Earth and Space Science Fellowship Program through grants NNX16AO28H and 80NSSC18K1114. We acknowledge support from NSF grants AST-1006676, AST-1126413, AST-1310885, AST-1517592, AST-1310875, AST-1907622, the NASA Astrobiology Institute (NAI; NNA09DA76A), and PSARC in our pursuit of precision radial velocities in the NIR. We acknowledge support from the Heising-Simons Foundation via grant 2017-0494 and 2019-1177. We acknowledge support from NSF grant AST-1909506 and the Research Corporation for precision photometric observations with diffuser-assisted photometry. Computations for this research were performed on the Pennsylvania State University’s Institute for Computational \& Data Sciences (ICDS). A portion of this work was enabled by support from the Mt Cuba Astronomical Foundation. RID is supported by NASA XRP 80NSSC18K0355 and the Alfred P. Sloan Foundation's Sloan Research Fellowship. Part of this research was carried out at the Jet Propulsion Laboratory, California Institute of Technology, under a contract with the National Aeronautics and Space Administration (NASA). GKS wishes to thank Kento Masuda for informative discussions on determining stellar inclinations from projected rotational velocities, stellar radii, and rotation periods.

These results are based on observations obtained with the Habitable-zone Planet Finder Spectrograph on the Hobby-Eberly Telescope. We thank the Resident astronomers and Telescope Operators at the HET for the skillful execution of our observations of our observations with HPF. The Hobby-Eberly Telescope is a joint project of the University of Texas at Austin, the Pennsylvania State University, Ludwig-Maximilians-Universität München, and Georg-August Universität Gottingen. The HET is named in honor of its principal benefactors, William P. Hobby and Robert E. Eberly. The HET collaboration acknowledges the support and resources from the Texas Advanced Computing Center. This is University of Texas Center for Planetary Systems Habitability Contribution 0001.

These results are based on observations obtained with the Apache Point Observatory 3.5-meter telescope which is owned and operated by the Astrophysical Research Consortium. We wish to thank the APO 3.5m telescope operators in their assistance in obtaining these data.

Based in part on observations at the Kitt Peak National Observatory, NSF’s NOIRLab (Prop. ID 0925-2018B; PI: G. Stefansson), managed by the Association of Universities for Research in Astronomy (AURA) under a cooperative agreement with the National Science Foundation. The WIYN 0.9m telescope is operated by WIYN Inc. on behalf of a Consortium of ten partner Universities and Organizations. WIYN is a joint facility of the University of Wisconsin–Madison, Indiana University, NSF’s NOIRLab, the Pennsylvania State University, Purdue University, University of California, Irvine, and the University of Missouri. The authors are honored to be permitted to conduct astronomical research on Iolkam Du’ag (Kitt Peak), a mountain with particular significance to the Tohono O’odham. We wish to dearly thank James Winsky at NSF's NOIRLab for his help to conduct some of the HDI observations.

This paper includes data collected by the \textit{Kepler} telescope. The \textit{Kepler} and \textit{K2} data presented in this paper were obtained from the Mikulski Archive for Space Telescopes (MAST). Space Telescope Science Institute is operated by the Association of Universities for Research in Astronomy, Inc., under NASA contract NAS5-26555. Support for MAST for non-HST data is provided by the NASA Office of Space Science via grant NNX09AF08G and by other grants and contracts. Funding for the \textit{K2} Mission is provided by the NASA Science Mission directorate. This research made use of the NASA Exoplanet Archive, which is operated by the California Institute of Technology, under contract with the National Aeronautics and Space Administration under the Exoplanet Exploration Program. This work has made use of data from the European Space Agency (ESA) mission {\it Gaia} (\url{https://www.cosmos.esa.int/gaia}), processed by the {\it Gaia} Data Processing and Analysis Consortium (DPAC, \url{https://www.cosmos.esa.int/web/gaia/dpac/consortium}). Funding for the DPAC has been provided by national institutions, in particular the institutions participating in the {\it Gaia} Multilateral Agreement.

\facilities{\textit{K2}, Gaia, HDI/WIYN 0.9m, ARCTIC/ARC 3.5m, HPF/HET 10m.} 
\software{AstroImageJ \citep{collins2017}, 
\texttt{astroplan} \citep{morris2018},
\texttt{astropy} \citep{astropy2013},
\texttt{astroquery} \citep{astroquery},
\texttt{barycorrpy} \citep{kanodia2018}, 
\texttt{batman} \citep{kreidberg2015},
\texttt{corner.py} \citep{dfm2016}, 
\texttt{celerite} \citep{Foreman-Mackey2017}, 
\texttt{dynesty} \citep{speagle2019}, 
\texttt{emcee} \citep{dfm2013},
\texttt{everest} \citep{luger2017}, 
\texttt{EXOFASTv2} \citep{eastman2017},
\texttt{HxRGproc} \citep{ninan2018},
\texttt{iDiffuse} \citep{stefansson2018b},
\texttt{Jupyter} \citep{jupyter2016},
\texttt{juliet} \citep{Espinoza2019},
\texttt{matplotlib} \citep{hunter2007},
\texttt{numpy} \citep{vanderwalt2011},
\texttt{MRExo} \citep{kanodia2019},
\texttt{pandas} \citep{pandas2010},
\texttt{PyAstronomy} \citep{czesla2019},
\texttt{pyde} \citep{pyde},
\texttt{radvel} \citep{fulton2018},
\texttt{SERVAL} \citep{zechmeister2018}.}

\bibliographystyle{yahapj}
\bibliography{references}

\newpage

\appendix

\section{Calcium Infrared Triplet Line Positions}
Table \ref{tab:cairt} lists the line index positions used to measure the Ca II IRT indices in the HPF spectra.

\begin{table}[H]
\centering
\caption{Line index positions used for Ca II IRT activity indicators. Wavelengths are given in vacuum wavelengths.}
\begin{tabular}{l l c}
\hline
Line \#        & Description            & Start \& End Wavelengths [\AA]\\\hline\hline
1              & Line Center            & [8499.930, 8500.780] \\
1              & Left Reference Region  & [8493.200, 8495.467] \\
1              & Right Reference Region & [8505.202, 8507.472] \\
2              & Line Center            & [8544.009, 8544.864] \\
2              & Left Reference Region  & [8535.887, 8538.737] \\
2              & Right Reference Region & [8551.562, 8554.412] \\
3              & Line Center            & [8664.086, 8664.953] \\
3              & Left Reference Region  & [8657.294, 8660.184] \\
3              & Right Reference Region & [8670.300, 8673.190] \\ \hline
\end{tabular}
\label{tab:cairt}
\end{table}

\section{Injection and Recovery Tests}
\label{sec:injectionappendix}
\begin{figure}[H]
\begin{center}
\includegraphics[width=0.7\textwidth]{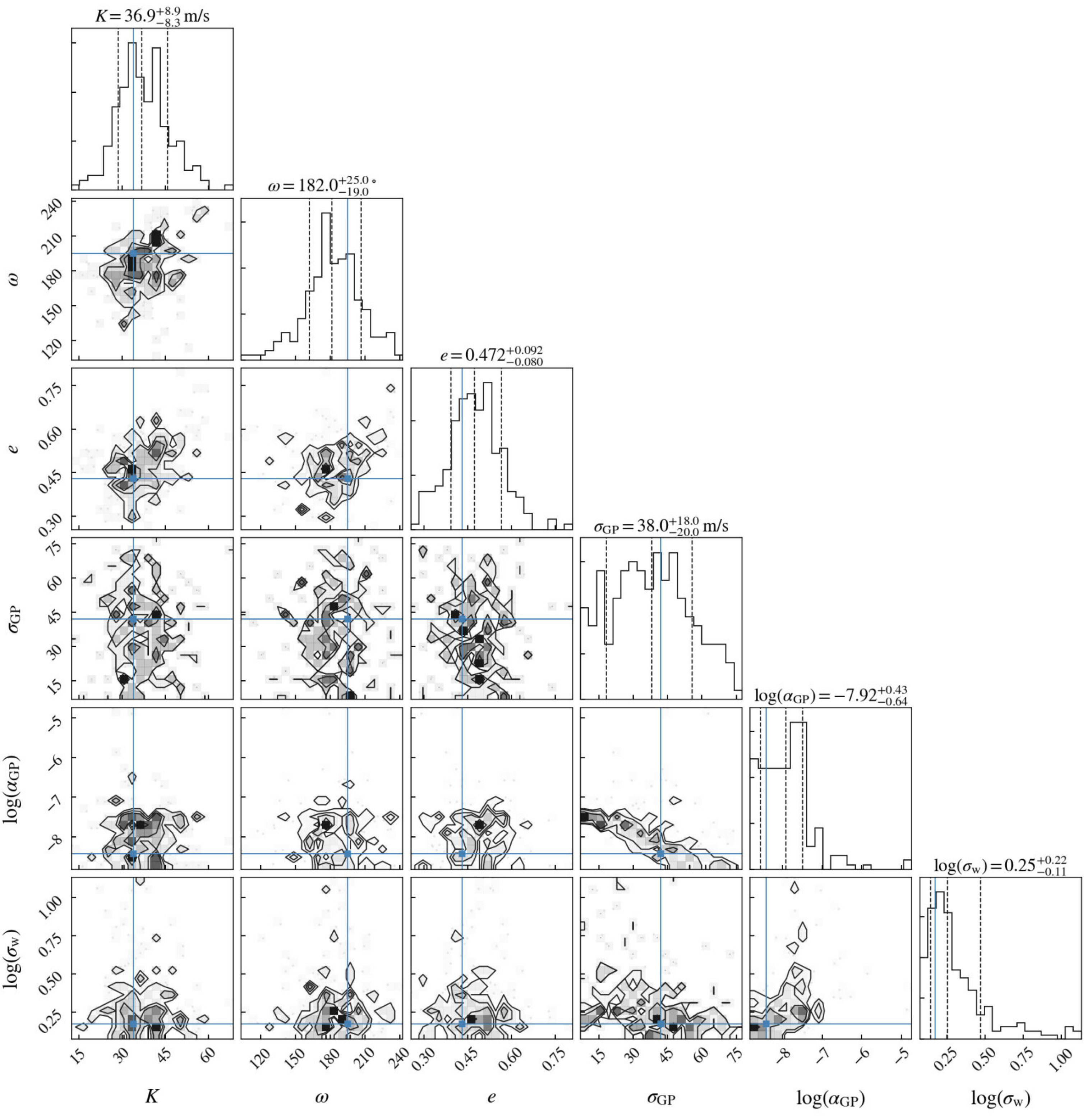}
\caption{Injection and Recovery Test I: results of 200 injection and recovery tests in synthetic HPF RV streams. The panels show the distribution of the means of individual posteriors. The known injected values are highlighted with the blue lines. The distribution of mean values agree well with the known values. Plot generated using \texttt{corner.py} \citep{dfm2016}.}
\label{fig:injection}
\end{center}
\end{figure}

\begin{figure}[H]
\begin{center}
\includegraphics[width=0.9\columnwidth]{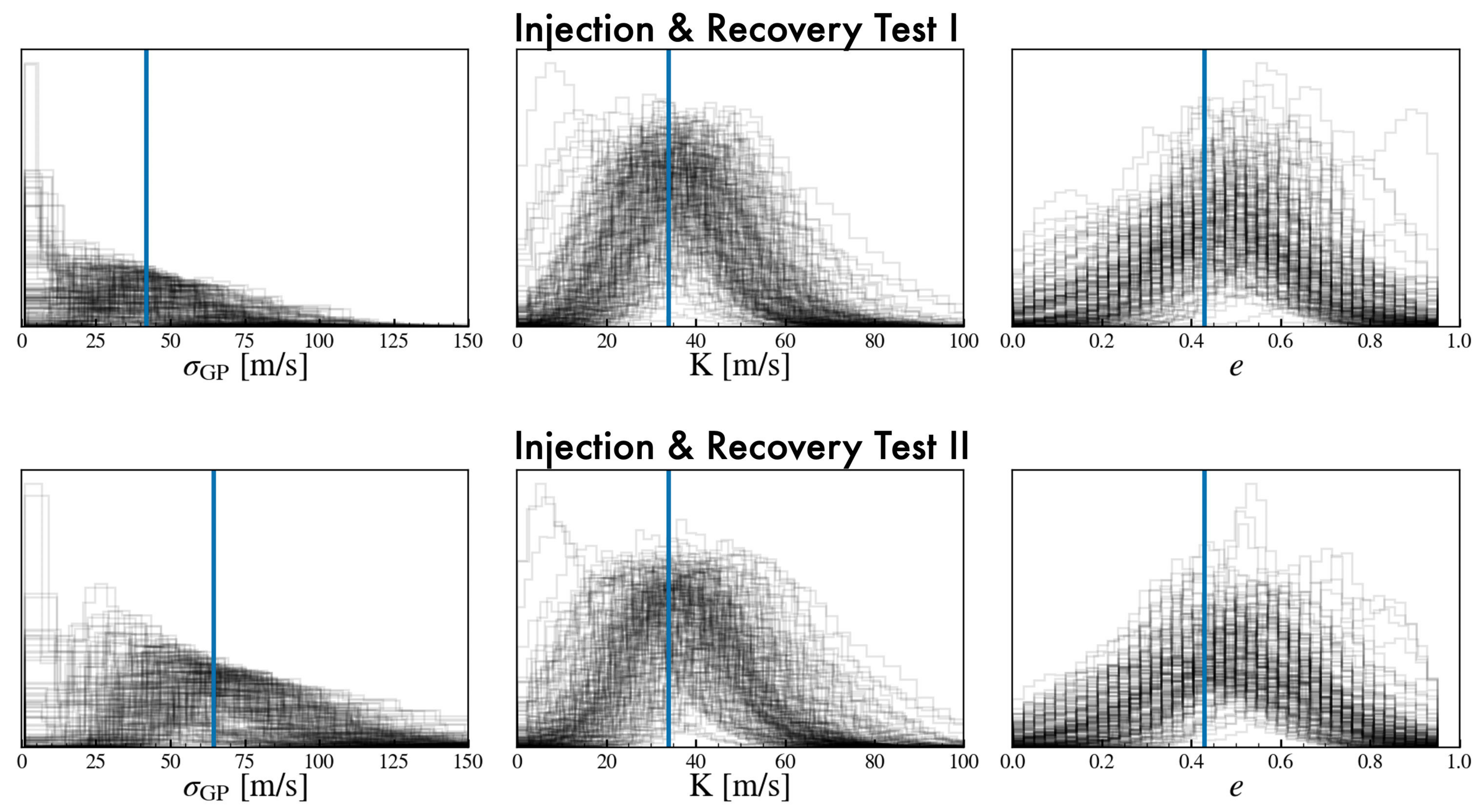}
\caption{Posteriors (black curves, 200 each) of the GP amplitude $\sigma_{\mathrm{GP}}$, the radial velocity semi-amplitude $K$, and eccentricity $e$, after two series of 200 injection and recovery tests in the HPF out-of-transit RVs. True values for $\sigma_{\mathrm{GP}}$, $K$, and $e$, are indicated by the blue lines. The first test (Test I; top panels) set $\sigma_{\mathrm{GP}}=42 \unit{m/s}$, $K = 34 \unit{m/s}$ and $e=0.43$ similar to the nominal median expected values from fit RV2. The true value of the semi-amplitude is within the 68\% and 95\% credible intervals in 80\% and 99.7\% of the cases, respectively. The second test (Test II; bottom panels) assumed that $\sigma_{\mathrm{GP}}=64.6 \unit{m/s}$, or at the 95th quantile value from fit RV2, while keeping the other parameters the same. Even with the higher assumed correlated noise, the true value of the semi-amplitude is still reliably recovered at a high coverage probability: the true value of the semi-amplitude is within the 68\% and 95\% credible intervals in 75\% and 97\% of the cases, respectively.}
\label{fig:posteriors}
\end{center}
\end{figure}

Table \ref{tab:injection}, and Figures \ref{fig:injection}, and \ref{fig:posteriors} summarize the results from our two series (Tests I and II) of synthetic injection and recovery tests. The methodology is further described in Section \ref{sec:injection}. Figures \ref{fig:injection}, and \ref{fig:posteriors}, show that the injected Keplerian parameters are consistently accurately recovered.

\begin{table}[H]
\centering
\caption{Summary of injected parameters for both series of tests considered. The results from the injection and recovery test are summarized in Figures \ref{fig:injection} and \ref{fig:posteriors}. $\mathcal{N}(m,\sigma)$ denotes a normal prior with mean $m$, and standard deviation $\sigma$; $\mathcal{U}(a,b)$ denotes a uniform prior with a start value $a$ and end value $b$, $\mathcal{J}(a,b)$ denotes a Jeffreys prior with a start value $a$ and end value $b$.}
\begin{tabular}{l l c c c}
\hline\hline
Parameter                     & Description                                        & Prior                              & Value: Test 1  & Value: Test 2         \\\hline
\multicolumn{5}{l}{\hspace{-0.2cm} Keplerian Parameters:}           \\                                                                                   
$P$ (days)                    &  Orbital period                                    & $\mathcal{N}(3.484548,0.000042)$   & \multicolumn{2}{c}{3.48456424}                \\ 
$T_C$                         &  Transit Midpoint - 2400000 $(\mathrm{BJD_{TDB}})$ & $\mathcal{U}(58515.63,58515.66)$   & \multicolumn{2}{c}{58515.642096}              \\ 
e                             &  Eccentricity                                      & $\mathcal{U}(0,0.95)$              & \multicolumn{2}{c}{0.43}                      \\ 
$\omega$                      &  Argument of Periastron                            & $\mathcal{U}(0,360)$               & \multicolumn{2}{c}{195}                       \\ 
$K$                           &  RV semi-amplitude ($\unit{m/s}$)                  & $\mathcal{U}(0,200)$               & \multicolumn{2}{c}{34}                        \\ 
\multicolumn{5}{l}{\hspace{-0.2cm} GP Hyperparameters:}           \\                                                                                                           
$P_{\mathrm{GP}}$             &  GP Period (days)                                  & $\mathcal{N}(1.8784,0.005)$        & \multicolumn{2}{c}{1.88178}                   \\ 
$\sigma_{\mathrm{GP}}$        &  RV GP Amplitude ($\unit{m/s}$)                    & $\mathcal{U}(10^{1},10^{5})$       & 42                                & 64.6    \\ 
$\Gamma_{\mathrm{GP}}$        &  Harmonic structure / scaling parameter            & $\mathcal{N}(8.0,1.9)$             & \multicolumn{2}{c}{7.4}                       \\ 
$\alpha_{\mathrm{GP}}$        &  Inverse length scale  ($\unit{days^{-2}}$)        & $\mathcal{J}(10^{-12},10^{-3})$    & \multicolumn{2}{c}{$3.8\times10^{-9}$}        \\ 
\multicolumn{5}{l}{\hspace{-0.2cm} Other HPF Parameters:}           \\                                                                                                        
$\sigma_{\mathrm{w}}$        &  HPF white noise RV jitter ($\unit{m/s}$)          & $\mathcal{J}(0.1,300)$             & \multicolumn{2}{c}{1.5}                       \\ 
$\mu_{\mathrm{HPF}}$          &  HPF RV offset ($\unit{m/s}$)                      & $\mathcal{U}(-200,200)$            & \multicolumn{2}{c}{0}                        \\ \hline
\end{tabular}
\label{tab:injection}
\end{table}

\section{HPF Radial Velocities}

\subsection{Out of transit}
Table \ref{tab:rvsout} lists the RVs from HPF used for the mass measurement of K2-25b, along with associated stellar activity indicators.

\begin{table}[H]
\centering
\caption{Out-of-transit RVs from HPF along with the differential line width (dLW), chromatic index (CRX), and the line indices for the three Ca II IRT triplet lines (Ca II IRT 1, 2 and 3), along with associated errors.}
\begin{tabular}{l c c c c c c}
\hline\hline
BJD &    RV [$\unit{m\:s^{-1}}$] &   dLW [$\unit{m^{2}\:s^{-2}}$]&   CRX  [$\unit{m\:s^{-1}\:Np^{-1}}$]&  Ca II IRT 1 &  Ca II IRT 2 &  Ca II IRT 3 \\
\hline
 2458424.95976 &    $29.8 \pm 15.6$ &    $288.7 \pm 74.2$ &     $88.1 \pm 233.7$ &  $0.878 \pm 0.005$ &  $0.790 \pm 0.006$ &  $0.662 \pm 0.005$ \\
 2458425.74905 &   $-90.6 \pm 35.6$ &   $312.2 \pm 170.0$ &  $-1649.2 \pm 478.6$ &  $1.097 \pm 0.014$ &  $0.931 \pm 0.017$ &  $0.828 \pm 0.013$ \\
 2458429.93716 &    $39.5 \pm 16.8$ &     $64.9 \pm 80.0$ &   $-445.4 \pm 209.2$ &  $0.930 \pm 0.006$ &  $0.826 \pm 0.007$ &  $0.703 \pm 0.005$ \\
 2458433.92567 &    $44.7 \pm 38.0$ &   $311.3 \pm 180.5$ &   $1049.5 \pm 630.4$ &  $0.854 \pm 0.012$ &  $0.764 \pm 0.016$ &  $0.685 \pm 0.012$ \\
 2458436.91691 &   $-56.8 \pm 23.9$ &  $-266.8 \pm 114.7$ &   $-145.4 \pm 332.6$ &  $0.975 \pm 0.008$ &  $0.871 \pm 0.010$ &  $0.708 \pm 0.008$ \\
 2458437.90840 &    $18.0 \pm 17.3$ &     $88.2 \pm 82.9$ &    $317.4 \pm 224.3$ &  $1.045 \pm 0.006$ &  $0.926 \pm 0.008$ &  $0.776 \pm 0.006$ \\
 2458439.91091 &    $-5.9 \pm 21.2$ &   $301.2 \pm 101.1$ &    $174.7 \pm 394.9$ &  $0.908 \pm 0.007$ &  $0.801 \pm 0.009$ &  $0.686 \pm 0.007$ \\
 2458441.89899 &    $62.8 \pm 22.3$ &   $176.8 \pm 106.5$ &   $-132.8 \pm 183.3$ &  $0.893 \pm 0.007$ &  $0.784 \pm 0.009$ &  $0.668 \pm 0.007$ \\
 2458442.89348 &   $-28.7 \pm 29.1$ &   $123.3 \pm 139.4$ &    $437.9 \pm 522.9$ &  $0.988 \pm 0.010$ &  $0.882 \pm 0.012$ &  $0.760 \pm 0.009$ \\
 2458443.71423 &    $38.9 \pm 24.4$ &   $296.1 \pm 116.1$ &     $19.8 \pm 352.8$ &  $0.908 \pm 0.007$ &  $0.810 \pm 0.009$ &  $0.712 \pm 0.007$ \\
 2458444.71135 &    $39.9 \pm 19.6$ &   $-315.5 \pm 94.6$ &     $34.1 \pm 181.4$ &  $0.946 \pm 0.006$ &  $0.844 \pm 0.008$ &  $0.720 \pm 0.006$ \\
 2458449.88697 &   $-80.4 \pm 22.3$ &   $169.7 \pm 107.3$ &   $-236.0 \pm 321.6$ &  $0.970 \pm 0.007$ &  $0.876 \pm 0.009$ &  $0.761 \pm 0.007$ \\
 2458451.68266 &   $-19.1 \pm 28.1$ &   $319.9 \pm 134.9$ &    $387.9 \pm 453.2$ &  $0.900 \pm 0.009$ &  $0.859 \pm 0.012$ &  $0.713 \pm 0.009$ \\
 2458473.80425 &    $-4.4 \pm 42.5$ &  $1752.3 \pm 196.0$ &    $178.1 \pm 669.2$ &  $0.893 \pm 0.015$ &  $0.762 \pm 0.019$ &  $0.644 \pm 0.014$ \\
 2458480.80404 &    $21.5 \pm 52.9$ &    $43.1 \pm 252.9$ &    $127.4 \pm 632.2$ &  $0.860 \pm 0.020$ &  $0.744 \pm 0.027$ &  $0.635 \pm 0.021$ \\
 2458487.78924 &  $-153.6 \pm 44.8$ &  $-523.5 \pm 218.1$ &    $137.8 \pm 710.1$ &  $0.891 \pm 0.017$ &  $0.718 \pm 0.024$ &  $0.640 \pm 0.017$ \\
 2458546.61407 &     $1.0 \pm 33.4$ &  $-390.2 \pm 162.4$ &    $434.9 \pm 462.0$ &  $0.895 \pm 0.010$ &  $0.829 \pm 0.013$ &  $0.677 \pm 0.010$ \\
 2458549.60481 &    $15.6 \pm 28.0$ &   $473.1 \pm 134.3$ &   $-178.2 \pm 444.0$ &  $0.885 \pm 0.008$ &  $0.825 \pm 0.011$ &  $0.684 \pm 0.008$ \\
 2458741.89127 &   $-36.4 \pm 37.6$ &  $-179.1 \pm 180.4$ &    $-55.6 \pm 587.0$ &  $0.907 \pm 0.012$ &  $0.831 \pm 0.017$ &  $0.708 \pm 0.012$ \\
 2458744.87566 &    $11.1 \pm 31.4$ &   $-48.3 \pm 149.9$ &  $-1098.4 \pm 405.1$ &  $1.040 \pm 0.011$ &  $0.960 \pm 0.014$ &  $0.823 \pm 0.010$ \\
 2458752.85672 &   $-43.7 \pm 31.5$ &   $393.2 \pm 149.1$ &   $-267.1 \pm 441.1$ &  $1.030 \pm 0.011$ &  $0.907 \pm 0.015$ &  $0.803 \pm 0.011$ \\
 2458804.90940 &  $-107.8 \pm 41.2$ &   $486.7 \pm 196.0$ &   $1583.8 \pm 570.3$ &  $0.902 \pm 0.014$ &  $0.801 \pm 0.018$ &  $0.667 \pm 0.014$ \\
 2458805.71647 &  $-117.9 \pm 28.5$ &   $433.8 \pm 135.7$ &  $-1673.7 \pm 264.1$ &  $0.874 \pm 0.010$ &  $0.753 \pm 0.013$ &  $0.674 \pm 0.010$ \\
 2458808.70580 &     $7.6 \pm 23.4$ &  $-300.2 \pm 112.8$ &     $64.8 \pm 464.9$ &  $0.946 \pm 0.008$ &  $0.889 \pm 0.011$ &  $0.726 \pm 0.008$ \\
 2458808.90127 &    $31.0 \pm 22.0$ &   $-61.7 \pm 105.8$ &    $-96.6 \pm 260.6$ &  $1.024 \pm 0.008$ &  $0.921 \pm 0.010$ &  $0.775 \pm 0.007$ \\
 2458811.88979 &    $28.6 \pm 30.4$ &  $-132.5 \pm 146.7$ &   $-412.6 \pm 421.3$ &  $0.918 \pm 0.010$ &  $0.859 \pm 0.013$ &  $0.708 \pm 0.010$ \\
 2458824.66447 &   $-97.2 \pm 44.0$ &   $722.8 \pm 209.6$ &    $247.7 \pm 747.0$ &  $0.907 \pm 0.015$ &  $0.797 \pm 0.020$ &  $0.677 \pm 0.015$ \\
 2458825.66903 &    $32.9 \pm 55.7$ &  $-289.9 \pm 271.4$ &    $365.9 \pm 792.5$ &  $0.939 \pm 0.018$ &  $0.878 \pm 0.024$ &  $0.757 \pm 0.018$ \\
 2458832.64612 &    $58.1 \pm 35.4$ &  $-230.3 \pm 172.1$ &    $738.2 \pm 645.5$ &  $0.899 \pm 0.011$ &  $0.792 \pm 0.015$ &  $0.692 \pm 0.012$ \\
 2458849.59344 &    $39.7 \pm 54.1$ &   $254.3 \pm 256.9$ &    $616.0 \pm 873.0$ &  $0.907 \pm 0.017$ &  $0.774 \pm 0.021$ &  $0.714 \pm 0.017$ \\
 2458849.77940 &    $41.6 \pm 32.7$ &  $-606.0 \pm 158.7$ &    $661.1 \pm 496.7$ &  $0.924 \pm 0.010$ &  $0.817 \pm 0.013$ &  $0.683 \pm 0.010$ \\
 2458852.77670 &   $-61.7 \pm 27.8$ &   $271.2 \pm 132.7$ &   $-400.7 \pm 457.1$ &  $0.901 \pm 0.008$ &  $0.821 \pm 0.011$ &  $0.675 \pm 0.008$ \\
\hline
\end{tabular}
\label{tab:rvsout}
\end{table}

\subsection{In transit}
Table \ref{tab:rvsin} lists the RVs from HPF used for the RM effect analysis. These RVs had an exposure time of 300s.
\begin{table}[H]
\centering
\caption{In-transit RVs from HPF.}
\begin{tabular}{l l l l l}
\hline\hline
Time ($\mathrm{BJD_{TDB}}$) & RV ($\unit{m/s}$) & RV Error ($\unit{m/s}$) & S/N & Transit \# \\ \hline
2458473.798377 &     21.502 &     92.473 &       31.5 & 1 \\
2458473.802296 &    -99.316 &     85.383 &       32.9 & 1 \\
2458473.806148 &     32.915 &     87.976 &       33.0 & 1 \\
2458473.810188 &     25.879 &     76.650 &       36.0 & 1 \\
2458473.814183 &     58.845 &     68.489 &       38.9 & 1 \\
2458473.818066 &     22.483 &     81.288 &       35.4 & 1 \\
2458473.822004 &    -29.955 &     70.497 &       38.1 & 1 \\
2458473.825933 &   -107.874 &     82.864 &       33.7 & 1 \\
2458473.829840 &     10.968 &    101.494 &       28.8 & 1 \\
2458473.833884 &      5.058 &    112.039 &       26.2 & 1 \\
2458473.837847 &     -5.540 &    111.578 &       26.1 & 1 \\
2458480.779096 &     93.172 &     81.592 &       32.5 & 2 \\
2458480.783031 &     84.061 &     74.764 &       35.2 & 2 \\
2458480.786978 &    289.359 &     70.453 &       36.9 & 2 \\
2458480.790926 &    118.139 &     59.718 &       42.5 & 2 \\
2458480.794850 &    147.978 &     61.526 &       41.0 & 2 \\
2458480.798729 &    -91.458 &     63.679 &       40.3 & 2 \\
2458480.802696 &    -89.542 &     72.040 &       36.7 & 2 \\
2458480.806659 &    -88.708 &     71.505 &       36.4 & 2 \\
2458480.810610 &     82.751 &     78.649 &       34.5 & 2 \\
2458480.814557 &    -30.039 &     93.241 &       29.1 & 2 \\
2458480.818470 &    -30.570 &    103.872 &       27.2 & 2 \\
2458487.763970 &    -32.264 &     92.523 &       29.1 & 3 \\
2458487.767653 &   -108.714 &     72.947 &       35.5 & 3 \\
2458487.771528 &   -163.954 &     69.931 &       37.3 & 3 \\
2458487.775525 &   -131.220 &     67.386 &       38.7 & 3 \\
2458487.779335 &    -91.287 &     66.037 &       38.8 & 3 \\
2458487.783313 &   -202.220 &     74.173 &       35.5 & 3 \\
2458487.787296 &   -203.813 &     88.859 &       30.1 & 3 \\
2458487.791204 &    -20.690 &     97.731 &       28.4 & 3 \\
2458487.795138 &   -138.737 &    108.448 &       25.3 & 3 \\
\hline
\end{tabular}
\label{tab:rvsin}
\end{table}

\end{document}